# ILACS-LGOT: A Multi-Layer Contrast Enhancement Approach for Palm-Vein Images

Kaveen Perera, Fouad Khelifi, Ammar Belatreche
Computer & Information Sciences department
University of Northumbria at Newcastle, United Kingdom

## Abstract

This article presents an extended author's version based on our previous work, where we introduced the Multiple Overlapping Tiles (MOT) method for palm vein image enhancement. To better reflect the specific operations involved, we rename MOT to ILACS-LGOT (Intensity-Limited Adaptive Contrast Stretching with Layered Gaussian-weighted Overlapping Tiles). This revised terminology more accurately represents the method's approach to contrast enhancement and blocky effect mitigation. Additionally, this article provides a more detailed analysis, including expanded evaluations, graphical representations, and sample-based comparisons, demonstrating the effectiveness of ILACS-LGOT over existing methods.

## 1. Introduction

Traditional authentication methods, such as PIN codes and passwords, have long been recognised as inefficient and insecure. In contrast, biometric recognition offers significantly more secure authentication by utilising one or more physiological traits [1]. Vascular biometric recognition systems identify individuals based on the unique patterns of blood vessels beneath the skin. These vascular networks are distinct for each person, even among identical twins. Moreover, these patterns remain relatively stable over time, enhancing the security and reliability of palm vein recognition systems. Additionally, palm vein images provide a higher degree of distinctive features with rich texture information, further strengthening their effectiveness.

Contactless palm vein recognition systems are non-invasive and hygienic, making them more user-friendly than other biometric recognition methods, which contributes to their higher user acceptance [2]. Near-Infrared (NIR) light penetrates the skin tissue, and the deoxygenated veins near the skin's surface absorb the NIR light, causing them to appear darker in captured images. However, skin tissue scatters NIR light, resulting in blurred vein patterns and low-contrast images. In combination with sensor noise from image capture devices, this makes feature extraction and subsequent image processing particularly challenging. As a result, a contrast enhancement pre-processing step is necessary before applying any feature extraction algorithm.

Several contrast enhancement methods have been explored in the literature to address the low contrast in palm-vein images. Common approaches include Linear/Nonlinear

Functional Enhancements (LNFE), Histogram Equalisation and Its Variants (HEs), and Hierarchical Enhancement Models (HEM) or Multilayer Enhancement (MLE-HE) [3].

LNFE techniques, such as intensity normalisation (IN) and gamma correction (GC), enhance visual quality but fail to incorporate neighbouring information. Consequently, they often struggle to sufficiently enhance the finer details essential for accurate biometric matching. HEs techniques, including Contrast Limited Adaptive Histogram Equalisation (CLAHE), aim to distribute intensity values more uniformly; however, they have been reported to perform poorly with the Scale-Invariant Feature Transform (SIFT) detector [4], [5]. While modifying CLAHE parameters can improve SIFT keypoint detection, such adjustments may introduce excessive noise and result in visible blocky artefacts between contextual regions.

HEM, which integrates multiple techniques such as Difference of Gaussians (DoG) and HE, offers enhanced contrast and detail preservation while performing well with SIFT-based feature detection. Nevertheless, these approaches may still omit crucial subspace information required for robust feature extraction. For example, the method proposed by Zhou and Kumar [6] demonstrated improvements in contrast but was affected by issues such as the blocky effect and the loss of detail in vein structures. Similarly, Local Binary Pattern (LBP)-based methods tend to remove essential subspace information, making them less suitable for applications requiring invariant feature extraction techniques.

In our previous work, we introduced the Multiple Overlapping Tiles (MOT) method [7] as a palm vein image enhancement technique. The proposed approach is a multi-layer hierarchical enhancement model (MLE-HE), which employs localised contrast stretching to enhance local contrast more effectively and a multi-layer method to eliminate the blocky effect.

This article builds upon that foundation, serving as an extended version of our earlier work. It offers additional insights, including enhanced experimental evaluations, graphical representations, and sample-based analyses, further validating the method's effectiveness. To align with the specific operations introduced in this extended study, we rename MOT to ILACS-LGOT (Intensity-Limited Adaptive Contrast Stretching with Layered Gaussian-weighted Overlapping Tiles). This new terminology better reflects the method's underlying principles, particularly its contrast enhancement mechanisms and approach to eliminating blocky artefacts.

The remainder of this article is structured as follows: Section 2 reviews related work on palm vein image contrast enhancement techniques, discussing their strengths and limitations. Section 3 provides a detailed explanation of the ILACS-LGOT method, covering its mathematical formulation, theoretical reasoning, and implementation steps. Section 4 describes the experimental setup, while Section 5 presents the findings and compares the results with existing methods followed by observations and recommendations. Finally, Section 6 concludes the article with key takeaways and future research directions.

## 2. Related Work

The most commonly used contrast enhancement technique in palm-vein recognition is CLAHE [8], [9], [10], [11], [12], [13], [14], [15], [16], [17]. Other studies have employed various multilayer HEM techniques to enhance the contrast of vein images.

Zhou and Kumar [18] estimated the background intensity profile by dividing the image into 32×32 tiles. To mitigate the blocky effect, these tiles overlap by 3 pixels. The average grey level of each block is then computed to estimate the corresponding background profile, which is resized to match the original image using bicubic interpolation before being subtracted from the original image. HE is subsequently applied to enhance the vein image.

Similarly, Lee [19] adopted a technique that divides the image into 16×16 non-overlapping tiles before computing the average grayscale values for each tile. However, unlike Zhou and Kumar [18], HE is not applied. Soh et al. [20] proposed an image enhancement process involving the application of CLAHE, followed by histogram stretching on 8×8 image blocks. A common limitation of these techniques is the blocky effect at the edges of the image tiles. Neither Soh et al. [20] nor Lee [19] address this issue, while the 3-pixel overlap used in Zhou and Kumar [18] is insufficient to fully resolve it. Additionally, HE can lead to the loss of finer details within the image.

Yang and Zhang [21] explored concepts from image dehazing techniques—which are designed to reduce the impact of scattered light in foggy images to improve background visibility—and adapted them to mitigate light scatter caused by skin tissues for vein image enhancement. Yang and Yang [22] introduced a multi-channel Gabor filter for vein pattern enhancement, where the filter parameters are automatically determined based on the width and orientation of the vein networks. Consequently, the four-orientation even-symmetric Gabor filter channels possess different centre frequencies, and the final enhanced image is obtained by fusing the four processed images.

Kim [23] proposed an image smoothing technique that considers the rate of intensity change. In regions where no vein patterns are present, the intensity change occurs at a slower rate than in areas containing vein structures. The difference between the original image and a smoothed version—termed the smoothing speed—is then used to distinguish vein patterns from the background. A common issue observed in the methods proposed by Yang and Zhang [21], Yang and Yang [22], and Kim [23] is that these image smoothing techniques discard critical subspace information, which is essential for detecting robust features.

Trabelsi et al. [24] calculated the global mean grey level and variance values before applying a convolutional kernel, which incorporates the global mean, variance, and an enhancement constant to compute the local response. However, this method introduces a darkening effect in areas where vein patterns branch out or are closely spaced, while also amplifying noise, necessitating additional filtering.

Van et al. [25] proposed an image enhancement method based on Local Binary Patterns (LBP). They refined Centre-Symmetric LBP (CS-LBP) to develop an improved variant known as Enhanced CS-LBP (ECS-LBP). Their enhancement method was compared against Gabor filters, LBP, and CS-LBP. However, LBP-based image enhancement methods tend to clip and discard most subspace information, making them unsuitable for invariant feature extraction techniques.

Yan et al. [26] explored an image sharpening technique, where the image is first multiplied by an enhancement factor of 10, then processed using the Prewitt high-pass filter. The resulting image is subsequently subtracted from the original image.

Kang et al. [27] proposed a method integrating Difference of Gaussian (DoG) filters with Histogram Equalisation (HE). A DoG filter with a 4:1 Gaussian kernel ratio is used as a band-pass filter to eliminate high frequencies, followed by the application of HE to the DoG-filtered image. This approach, known as DoG-HE, shares the same limitations as previously discussed methods, as it also discards subspace information, making it unsuitable for robust feature extraction.

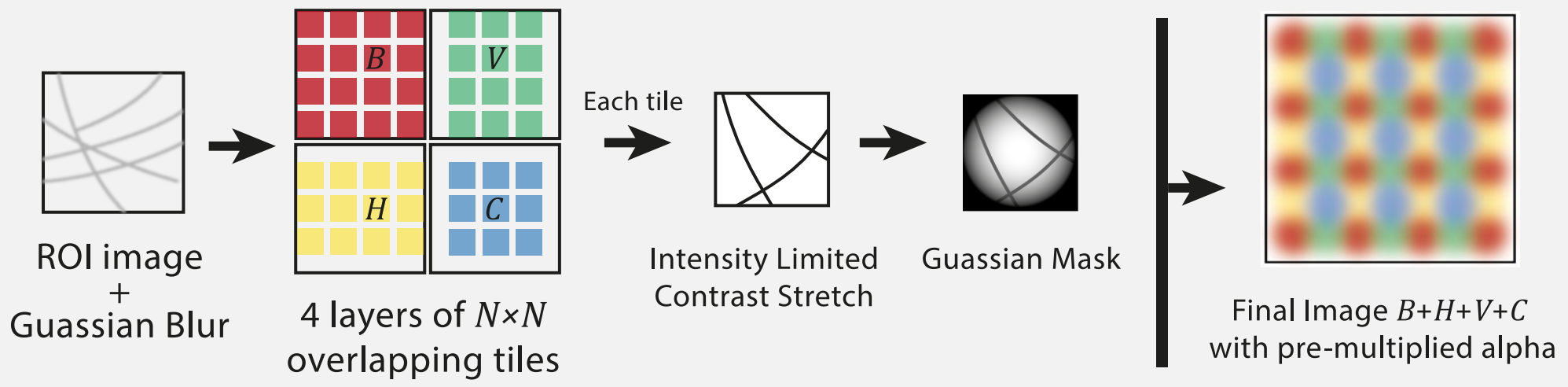

**Figure 1:** The flow graph of the ILACS-LGOT method.
(Gaps between the tiles added for visual clarity and illustration purposes only).

## 3. The Proposed Method

The proposed method consists of two components: a localised contrast enhancement operation and a technique to mitigate the blocky effect.

This method divides the input image into contextual regions, referred to as image tiles. Each tile then undergoes a modified contrast stretching operation, which incorporates a clipping control mechanism known as Intensity-Limited Adaptive Contrast Stretching (ILACS).

To address the blocky effect, the method employs multiple overlapping layers of image tiles. Each layer targets different sections of tile edges and vertices, effectively blending transitions between adjacent tiles. This approach is termed Layered Gaussian-weighted Overlapping Tiles (LGOT).

The final intensity value of a pixel is computed as a weighted sum of intensity values from all overlapping tiles containing that pixel, with weights assigned using a Gaussian mask. The combined ILACS-LGOT approach ensures a seamless transition across tile boundaries, effectively eliminating the blocky effect commonly associated with local contrast enhancement techniques. **Figure 1** illustrates the flow graphs for ILACS-LGOT method. The figures depict the sequence of operations involved in processing the image tiles and blending their overlapping regions. The following subsections provide detailed explanations of both components of the proposed method

### 3.1. Intensity Limited Adaptive Contrast Stretch (ILACS)

The underlying concept of the proposed method is a localised contrast stretch operation referred to as Adaptive Contrast Stretch (ACS). This method enhances the contrast of palm-vein images by dividing the image into square tiles of size $N \times N$ and stretching the contrast of each tile within the available dynamic range using ***Equation (1)*** [28]. The output is scaled back to the 0–255 range for the 8-bit palm-vein images used in this study by multiplying with 255.

Following are the steps taken to generate an ACS image:

- Subdivide the image into square tiles.
- Find $\min(T_N)$, the minimum pixel values of the particular $T_N$ tile.
- Find $\max(T_N)$, the maximum pixel values of the particular $T_N$ tile.
- Apply the $F(T_{N_i})$ function in ***Equation (1)*** for each pixel denoted as $T_{Ni}$.
- Repeat all the above steps for all $T_N$ tiles.

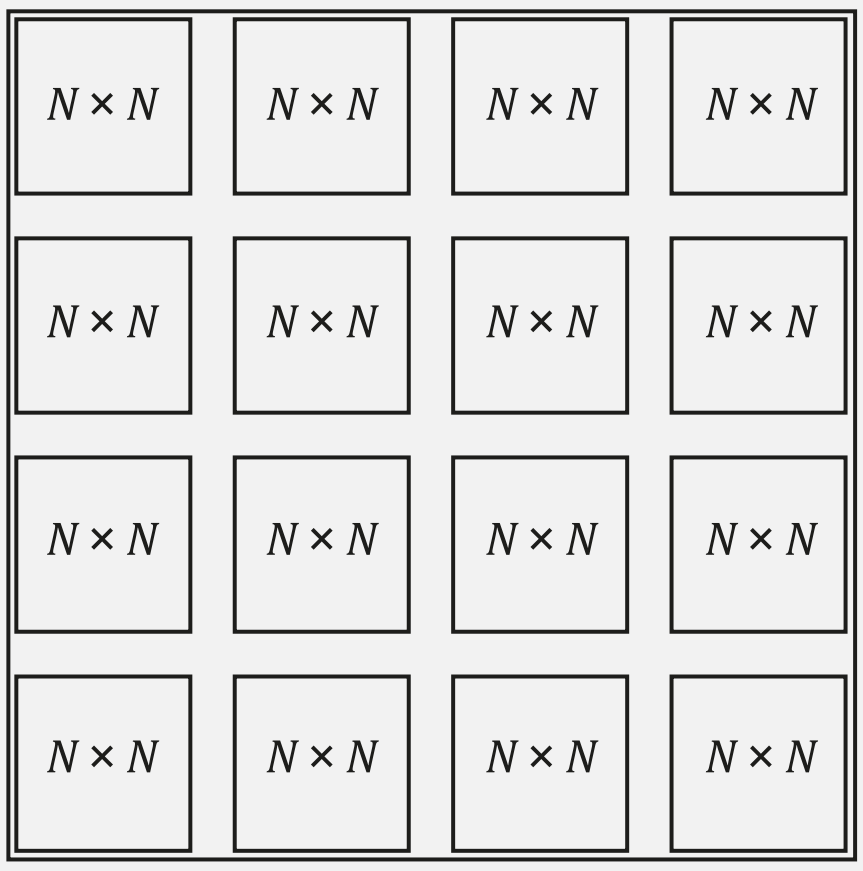

**Figure 2:** Image tiles of size $N$×$N$. ***Equation (1)*** is applied to every tile to stretch the contrast within the available dynamic range.
(Gaps between the tiles added for visual clarity and illustration purposes only).

$$F(T_{N_i}) = \left(\frac{T_{N_i} - \min(T_N)}{\max(T_N) - \min(T_N)}\right) \times 255 \qquad (1)$$

Where:

$N$ = length of a tile in pixels,

$T_N = N \times N$ sized image tile,

$i$ refers to an individual pixel in the image tile $T_N$.

$F(T_{N_i})$ in ***Equation (1)*** is a point operation. $T_N$ represents a subdivided square tile, denoted as an $N \times N$ matrix. The min( ) and max( ) functions calculate the minimum and maximum pixel values of the respective tile, and $T_{N_i}$ refers to the $i^{th}$ pixel of such a tile. The function is applied to each pixel of the $T_N$ tile to compute its new value.

If parts of the image have a uniform colour, such as a dark background in a palm-vein image, a binary matte can be used to mask these regions effectively to mitigate the division-by-zero error when the denominator of ***Equation (1)*** reaches zero. **Figure 2** provides an illustration of the subdivided tiles, while **Figure 3** presents the histograms of a sample tile before and after applying contrast stretch.

Contrast stretching within the available dynamic range can be susceptible to noise and artefacts. To address this issue, it is suggested in [28] to scale down the output of ***Equation (1)***. Adaptive Histogram Equalisation (AHE) suffers from a similar issue, as it can amplify noise in nearly uniform regions of an image [29]. To address this, the clipping mechanism in CLAHE sets a threshold, or a limit, for the maximum number of pixels in any histogram bin of each tile. When a bin exceeds this threshold, the excess pixels are redistributed uniformly across all bins to prevent noise amplification. In practice, a global clipping value is often used consistently across all contextual regions to maintain uniform contrast enhancement throughout the image [30].

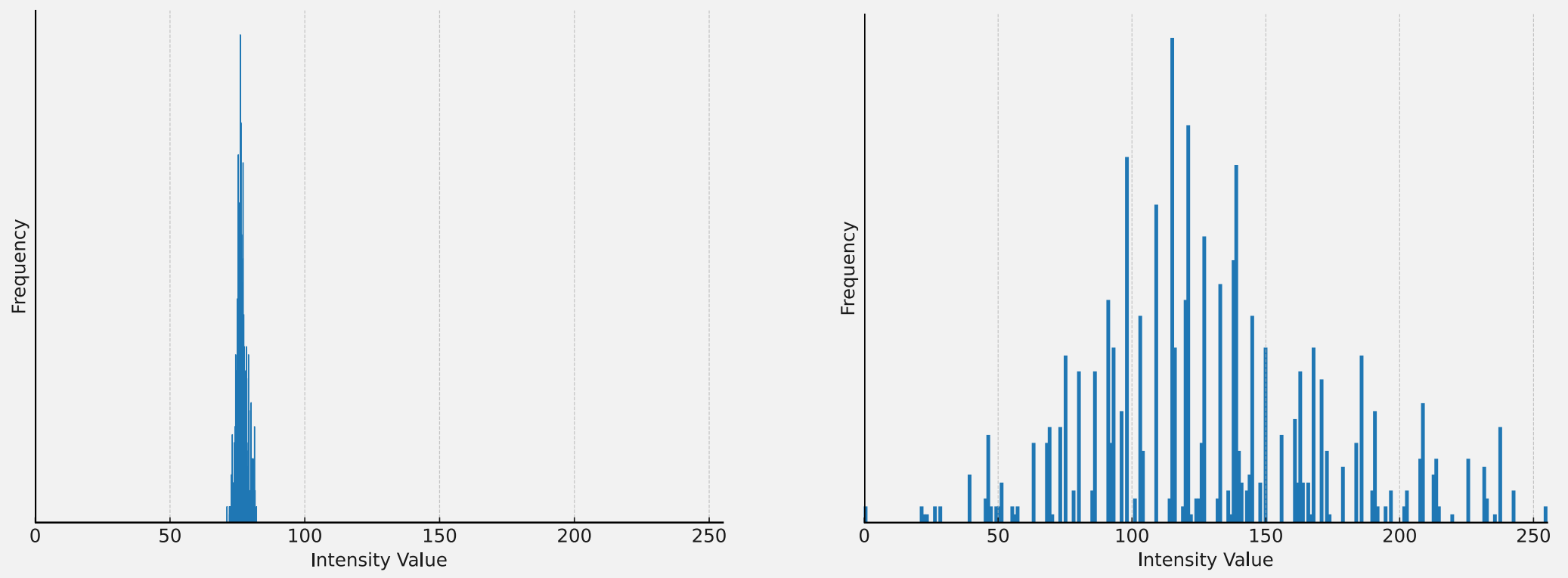

[a] [b]

**Figure 3:** Histograms of a sample tile,
[a] - before and [b] - after applying Contrast Stretch.

$$F(T_{N_i}) = \left(T_{N_i} - \min(T_N)\right) \times \left\lfloor \frac{255}{\max(T_N) - \min(T_N)} \right\rfloor \quad (2)$$

The symbol $\lfloor \ \rfloor$ in ***Equation (2)*** indicates the floor function.
The floor function, $\lfloor x \rfloor$, is defined as the greatest integer $n$ such that $n \leq x$:
$\lfloor x \rfloor = max\{n \in \mathbb{Z} \mid n \leq x\}$

In ACS, scaling down the contrast stretch using a fixed percentage as suggested by [28] will mimic a similar global functionality to that of CLAHE. However, the local contrast distribution of palm images can vary; hence a global operation is not suitable. While maintaining a truly localised operation, the denominator of ***Equation (1)*** was combined with the multiplier and applied using a floor function to scale down the contrast stretch. This effectively acts as a step function, which automatically sets the scaling down based on the local pixel intensities of a given image tile. This method is referred to as Intensity Limited Adaptive Contrast Stretch (ILACS). The modified formula of the ILACS method is presented in ***Equation (2)***.

ILACS is a point operation which does not consider the intensities of the neighbouring tiles. As a result, two adjacent pixels with minimal intensity difference between them but reside in two different tiles, can be remapped to very different intensities based solely on the minimum and maximum pixel values of the respective tile, producing a sharp edge between the tiles. **Figure 4** [a] and [b] demonstrate this blocky artefact that occurs at the edges of each tile. This is a well-known problem in localised image operations, where methods must find a balance between local enhancement and global consistency.

## 3.2. Layered Gaussian-weighted Overlapping Tiles (LGOT)

To mitigate the blocky effect, AHE utilises a method that incorporates several sampling techniques to smooth transitions between tiles. In AHE, sampling can be done at intervals equal to the contextual region size (mosaic sampling), half the contextual region size (half-sampling), or every pixel (full-sampling). Variations include weighted AHE, which assigns more weight to pixels near the centre of the contextual region [31]. In their study, Zhou and Kumar [6] employed a technique where they used 32×32 image tiles overlapped by 3 pixels. However, a 3-pixel overlap is insufficient to effectively surpass the blocky effect.

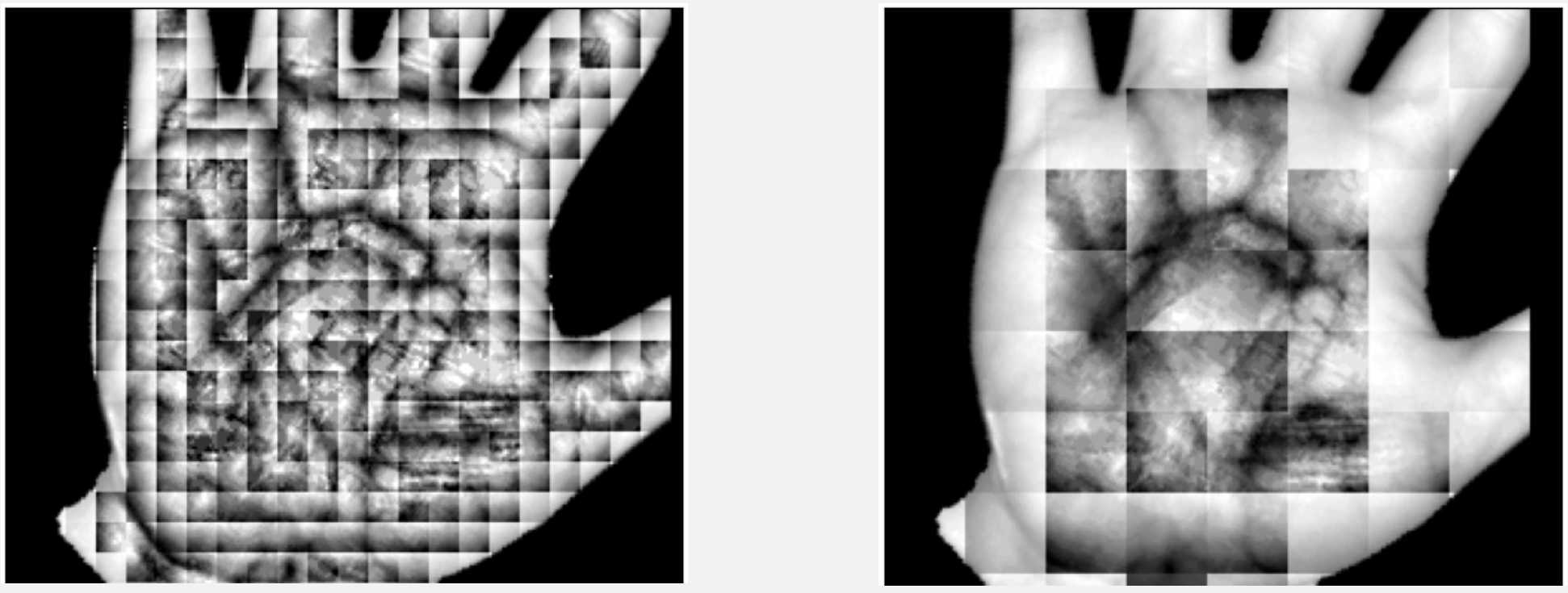

[a]: $N = 12$ (Tile size - 12×12) [b]: $N = 32$ (Tile size - 32×32)

**Figure 4:** Blocky effect at the edges of each tile.

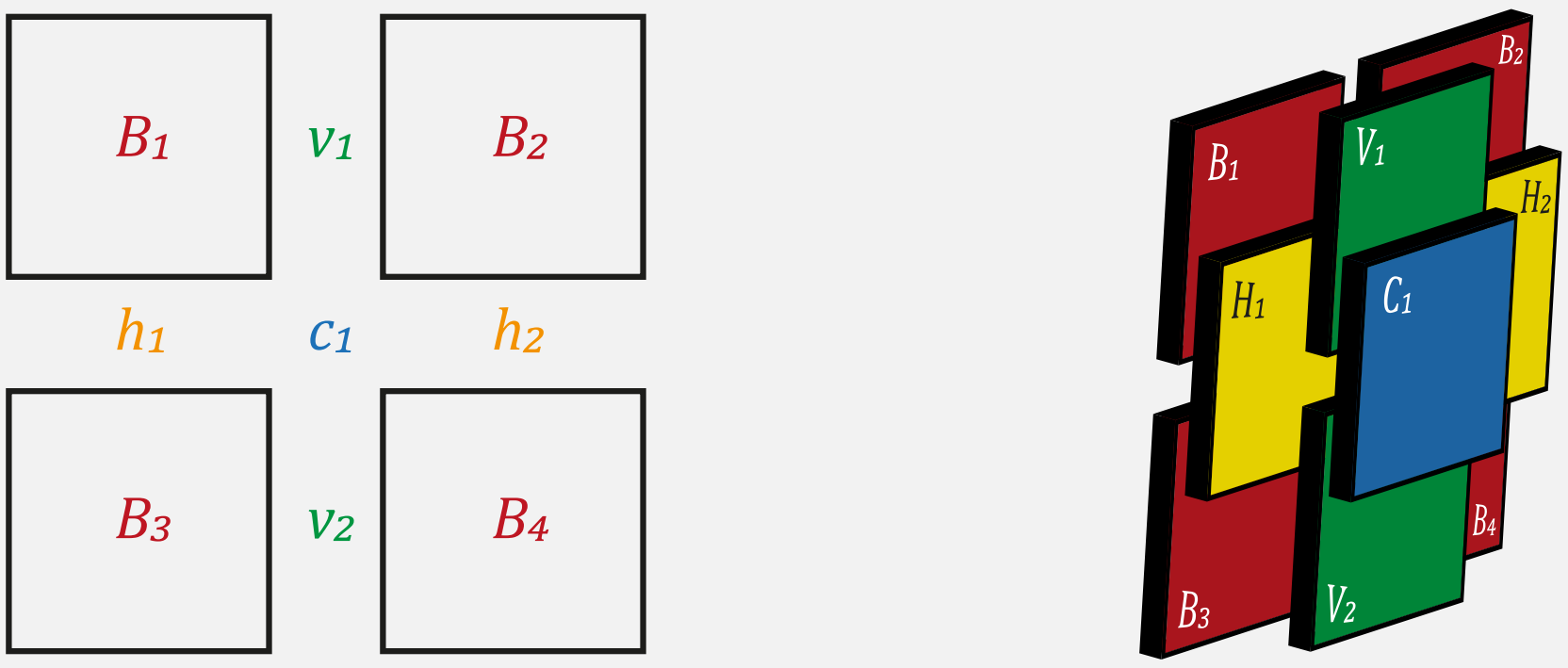

[a]: Edges and the common vertex. [b]: Overlapping tiles at the edges and the common vertex.

**Figure 5:** Examining 4 neighbouring image tiles.

Building upon the strengths of AHE, the Layered Gaussian-weighted Overlapping Tiles (LGOT) method further refines this approach by incorporating the concept of half-sampling across multiple layers. This method uses Gaussian-weighted pixel values for each image tile, effectively mitigating the blocky effect while enhancing contrast.

**Figure 5** [a] illustrates four neighbouring image tiles ($B_1$ – $B_4$) from the base image layer (the original ROI image enhanced with ILACS shown in **Figure 4**). $v_1$ and $v_2$ represent common vertical edges, while $h_1$ and $h_2$ represent common horizontal edges. $C_1$ is the common vertex across all four tiles. The ILACS-LGOT method uses multiple overlapping tile layers at these edges and vertices. In **Figure 5** [b], overlapping image tiles ($H_1$ – $H_2$ and $V_1$ – $V_2$) are generated at these edges and the common vertex ($C_1$). Three overlapping layers were created to cover the entire image.

**Figure 6** illustrates the four layers of overlapping tiles, named *B*, *H*, *V*, and *C*. The second and third layers *(H, V)* centre on the horizontal and vertical edges of a sample image with a 4×4 base tile layer *(B)*, while the fourth layer *(C)* centres on the common vertices. In this example, a total of 49 ILACS tiles were used.

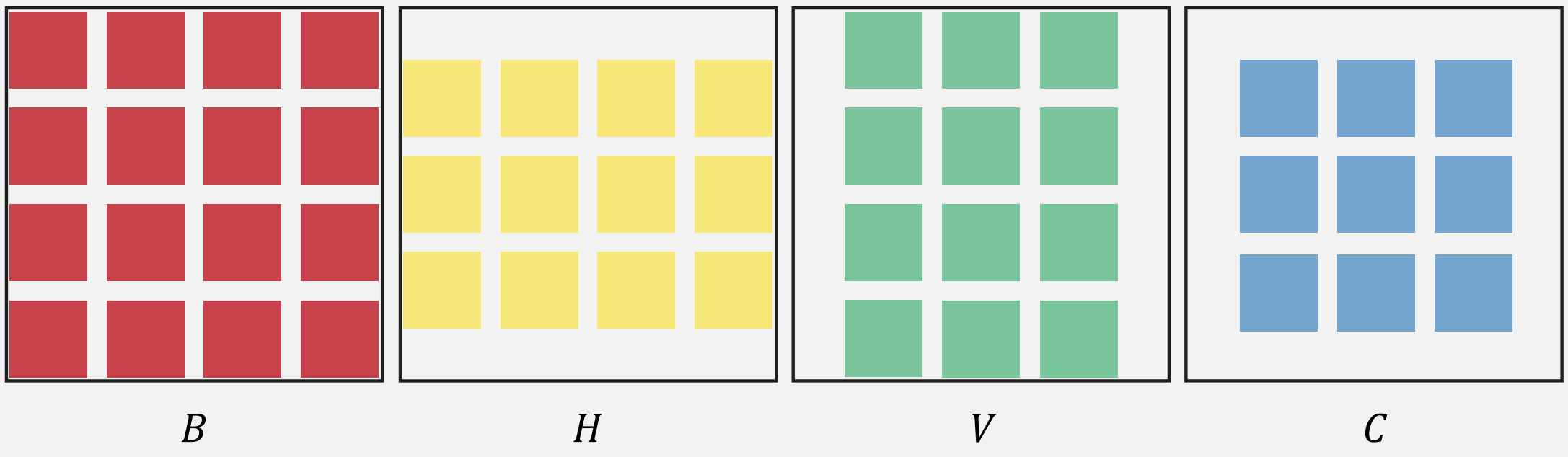

**Figure 6:** Overlapping tile layers.
*B* - Base tiles layer, *H* - Horizontal overlaps,
*V* - Vertical overlaps, *C* - Common vertex overlaps.
(Gaps between the tiles are added for visual clarity and illustrative purposes only).

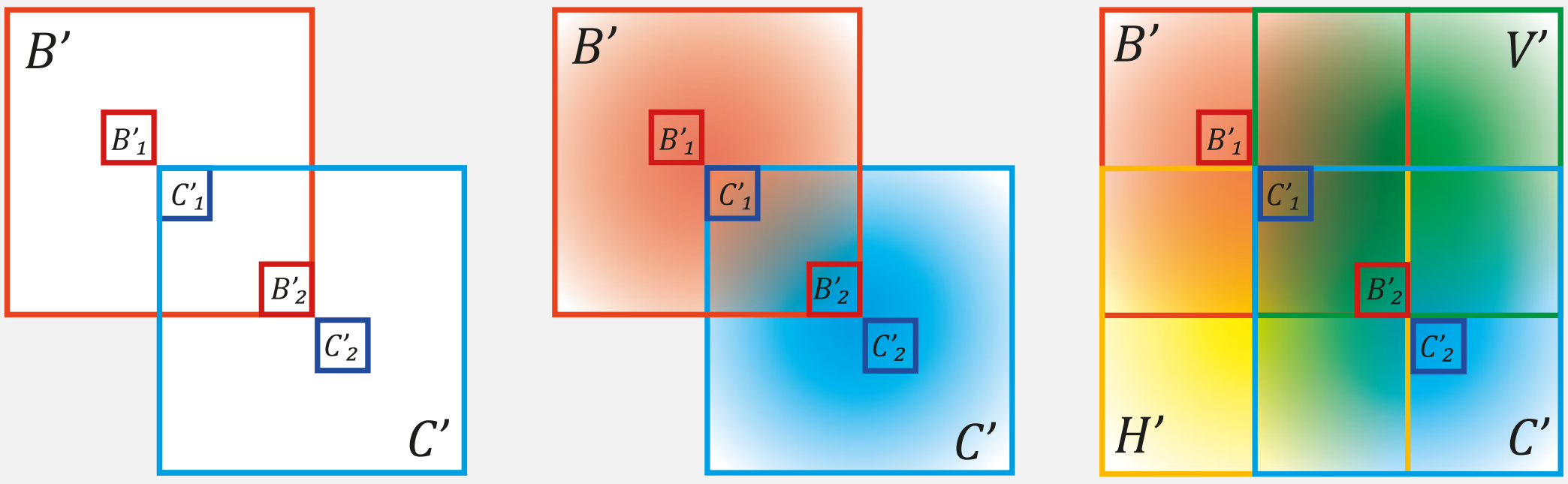

[a]: No radial decay applied. [b]: Radial decay on 2 overlapping tiles. [c]: Radial decay on 4 overlapping tiles.

**Figure 7:** Examining and supressing blocky effect with a radial decay mask. The smooth transitions may not appear accurately on some computer screens or in printed media.

The primary cause of the blocky effect is that two neighbouring pixels with very similar values are remapped into vastly different values when they are located in two distinct image tiles.

Consider the example scenario in subfigures of **Figure 7**, where $B'$, $H'$, $V'$, and $C'$ are individual tiles from the corresponding $B$, $H$, $V$, and $C$ tile layers depicted in **Figure 6**. In **Figure 7** [a], $B'$ and $C'$ are two overlapping tiles. $B'_1$, $C'_1$ and $B'_2$, $C'_2$ are adjacent pixel pairs which reside in the centre of the $B'$ and $C'$ tiles, respectively.

The value of the $B'_1$ pixel is not influenced by tile $C'$, and the value of the $B'_2$ pixel is not influenced by tile $B'$. $C'_1$ and $B'_2$ are affected by both overlapping tiles.

As the ILACS method considers all the pixels in each image tile, the value of $B'_2$ is still directly affected by the values of $B'_1$ and $C'_1$, and the value of $C'_1$ is directly affected by the values of $C'_2$ and $B'_2$. $C'_1$, $C'_2$ and $B'_1$, $B'_2$ are not adjacent pixel pairs; hence, when producing ILACS image tiles, their influence towards each other ($B'_1 \leftrightarrow B'_2$, $C'_1 \leftrightarrow C'_2$) should be minimal to none. This can be achieved using a weighted average for each pixel on overlapping tiles based on the location of the pixel within each tile. The centre pixels of each tile should have a weighting of 1, which gradually reduces to 0 at the edges. This hypothesis can be visually demonstrated using a radial decay effect, as illustrated in **Figure 7** [b]. It can be observed that the influence exerted on $C'_1$ and $B'_2$ pixel pairs by the pixels at the edges of the respective tiles is minimised.

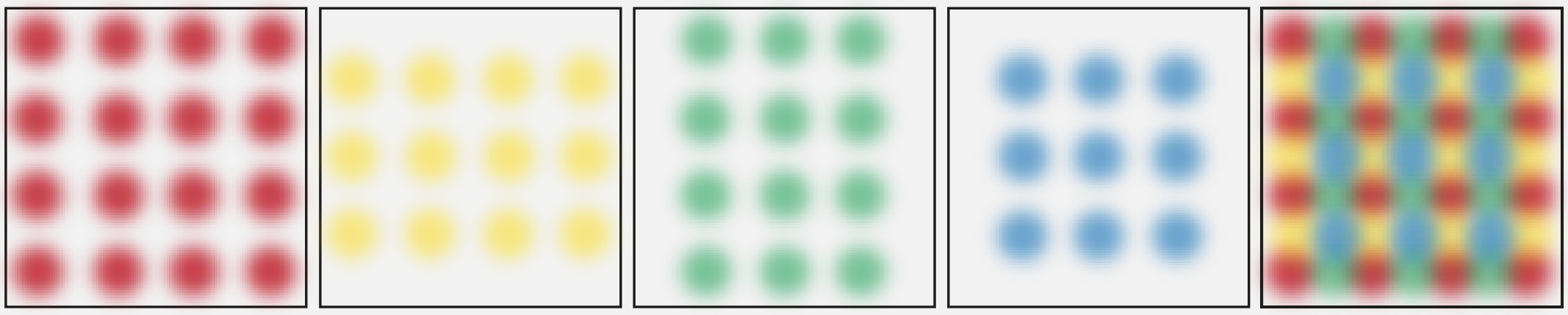

[a]: $B$ [b]: $H$ [c]: $V$ [d]: $C$ [e]: $B+H+V+C$

**Figure 8:** Overlapping tile layers $B$, $H$, $V$ and $C$ after applying a radial decay effect. The overlapping layers are combined to produce the final ILACS-LGOT image.

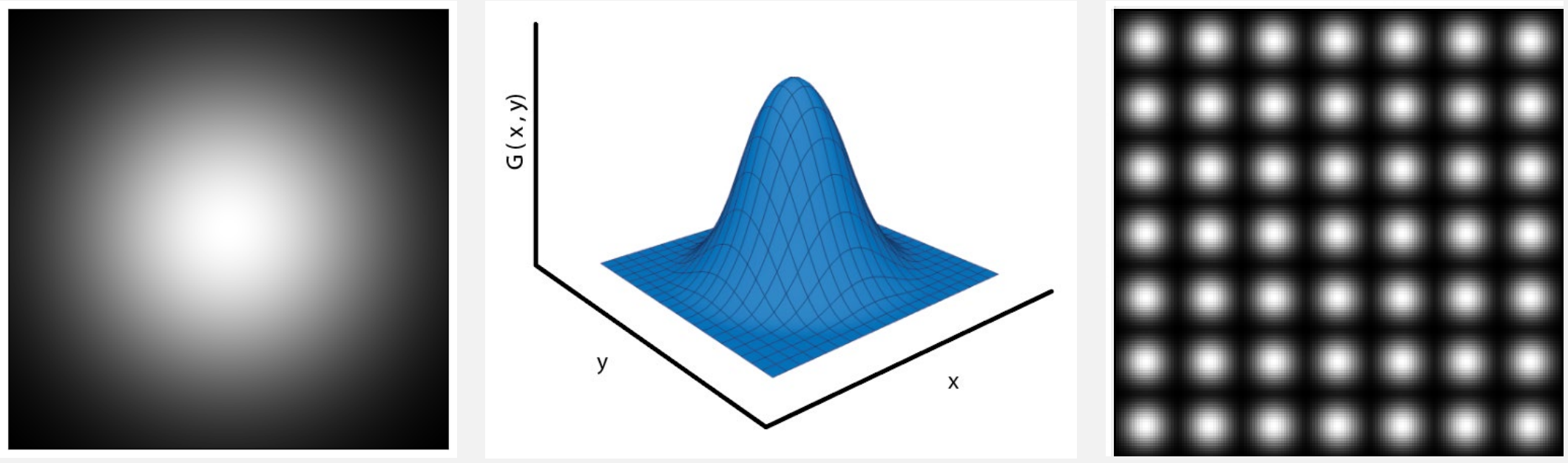

[a] [b] [c]

**Figure 9:**

[a] - Visualization of a radial gradient generated using the 2D isotropic Gaussian distribution with σ=1.

[b] - 3D representation of the Gaussian distribution.

[c] - A grid of radial gaussians distributions used as an alpha channel.

**Figure 7** [c] further expands this demonstration with four colour-coded tiles $B'$, $H'$, $V'$, and $C'$, to represent the four edge and vertex overlapping layers discussed above and presented in **Figure 6** as $B'$, $V'$, $H'$, and $C'$. Edge outlines on the tiles are retained to emphasise the overlapping sections between the tiles. The smoothness of the blending can be controlled by scaling up the radial decay rate on each tile.

In **Figure 7** [b] and [c], some colours may appear bright and prominent due to how the human visual cortex responds to brightness, saturation levels, and different hues of the colour spectrum. Furthermore, the smooth transitions may not appear accurately on some computer screens or in printed media. **Figure 8** illustrates the four layers, $B$, $H$, $V$, and $C$, from **Figure 6**, applied with weights using a radial decay effect onto each tile, and the ILACS-LGOT image produced by combining the four layers.

The radial decay effect can be achieved by utilising a radial gradient, which can be generated using a two-dimensional (2D) isotropic Gaussian function, as per the formula in ***Equation (3)*** [32]. The 2D isotropic Gaussian is derived from a mathematical function and is thus independent of any image or dataset. In ***Equation (3)***, $x$ and $y$ are the width and height of an image tile, and $\sigma$ is the standard deviation.

$$G(x,y) = \frac{1}{2\pi\sigma^2} e^{-\frac{x^2+y^2}{2\sigma^2}} \quad (3)$$

**Figure 9** [a] visualises a radial gradient generated using ***Equation (3)***. The image is rendered from the normalised (to 8-bit) matrix

generated with $G(x, y)$ (function values were generated between 0: black – 1: white). **Figure 9** [b] is a 3D representation of the same radial gradient. The slope (or the falloff) of the Gaussian distribution can be adjusted with $\sigma$ in ***Equation (3)***, to scale up or down the radial decay effect. **Figure 9** [c] is a sample grid of radial gradients generated to serve as weights for the base tile layer.

$$I_N = [(1 - \alpha)\, b] + \alpha\, f \qquad (4)$$

To understand how these weights apply to multiple layers and the process of producing the final ILACS-LGOT image, the weights can be utilised as a transparency or alpha channel ($\alpha$) for each layer [33]. ***Equation (4)*** is used to blend an image $f$ using a greyscale alpha channel α over a background image $b$ to create a new image $I_N$. This is also referred to as the over-operator [34]. This formula assumes the background to be a solid image without an applied alpha channel.

When the background is black and is equal to 0, the over operator can be reformulated as below, with α now acting exclusively as a weighting to the foreground image.

$$I_P = \alpha\, f \qquad (5)$$

***Equation (5)*** can be applied to any image with an associated alpha channel and is referred to as pre-multiplying in computer graphics and digital image compositing [35]. Hence, $I_P$ is used in ***Equation (5)*** to refer to a pre-multiplied image. The final image is then produced by taking the sum of all the weights pre-multiplied image layers. The sum of weights from every layer on a given pixel should always equal 1.

## 4. Evaluation Methodology

The evaluation of the ILACS-LGOT method is presented in two stages. In the first stage, the contrast enhancement results are presented, along with their respective histograms and the outcomes of the SIFT feature detector. This stage also includes an evaluation of SIFT keypoint matching and parameter analysis.

The second stage involves a performance comparison using EER as the measure of performance, where the ILACS-LGOT method is used as the contrast enhancement method in existing palm-vein recognition systems. Finally, recommendations for using the ILACS-LGOT methods are provided. This section will detail the experimental setup: the dataset, the ROI, and the feature matching and filtering techniques used for performance comparison.

For this study, the 850nm NIR subset of the publicly available CASIA dataset [36] was utilised. The CASIA Multi-Spectral dataset consists of six images per left and right hand, captured under different orientations and varying levels of finger and thumb stretch from 100 participants. The images were acquired under six different wavelengths: 460nm, 630nm, 700nm, 850nm, 940nm, and white light. Each wavelength includes a total of 1,200 images, all stored in JPEG format. As a result, the dataset contains image sensor noise and artifacts introduced by lossy JPEG compression, which are present in all images.

The dataset was divided into 20% (240 images) for testing during the development phase and 80% (960 images) for evaluation. To maximise the number of subjects, the left and right hands of a single subject were treated as two separate subjects, except that the opposite palms of a particular subject were not matched against each other.

Many previous studies which attempted palm-vein identification limited their algorithm to a smaller ROI [37], [38], [39], [40]. The ROI is focused on the central palmar region (centre part) of the palm, yet through visual observation of images from available datasets, it was observed that most of the palm veins are located towards the thenar eminence (from the base of the thumb to the wrist), hypothenar eminence (from the base of the little finger to the wrist), and upper palmar regions (refer to **Figure 10** [a]).

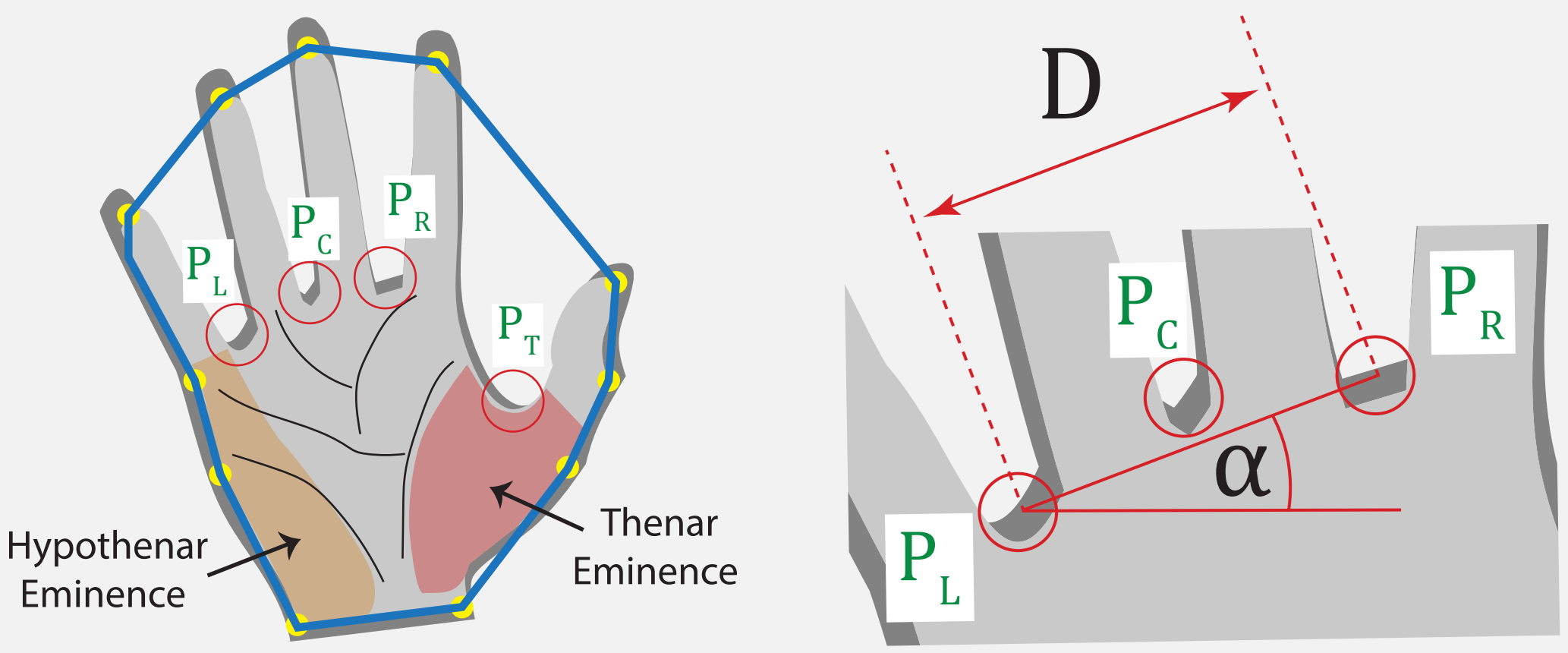

[a] [b]

**Figure 10:**

[a] - Palm valley point detection method as described in [136]. The image is first applied with OTSU thresholding and Suzuki contour detection. Then Sklansky's convex hull algorithm is used to identify the fingertips and the valley points.
$P_L$, $P_C$, $P_R$, and $P_T$ refers to left, centre, right, and thumb valley points, respectively.

[b] - Calculating the rotation angle 'α' and the distance 'D' between the left and right valley points.

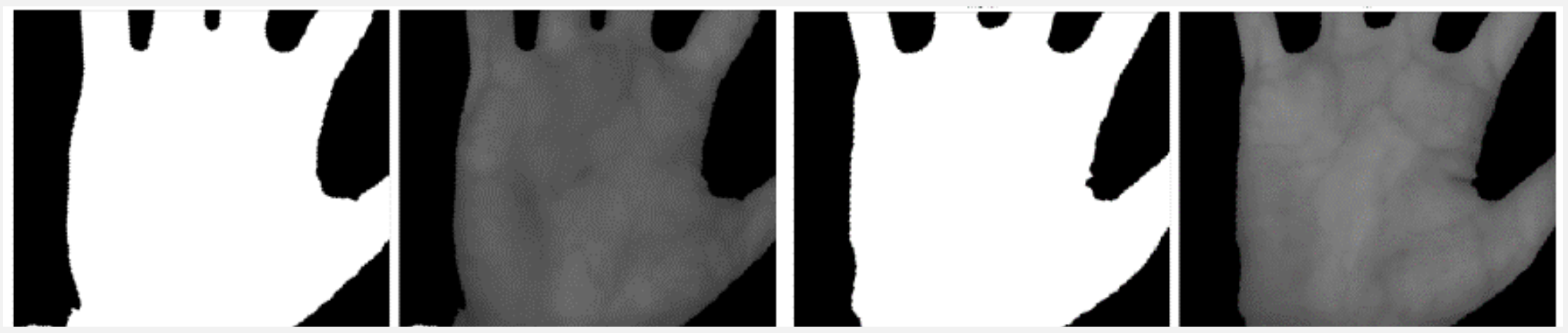

[a] [b]

**Figure 11:** Extracted ROI images and the corresponding mattes/masks.

To maximise the usable image data, the entire palm was used instead of a limited ROI. Similar methods were used by [41], [42], [43]. The valley points of the palm images were identified using the system described in [38] (refer to **Figure 10** [a]). Their method identifies the thumb, fingertips and valleys of the hand by applying Sklansky's convex hull algorithm along with OTSU thresholding and Suzuki contour detection. This also produced a binary matte of the palm.

Then the angle between the left and right valley points situated between the fingers was calculated, and the image was rotated to normalise the angle between these valley points to the horizontal axis (refer to **Figure 10** [b]). The distance between the left and right valley points, denoted as 'D' in **Figure 10** [b] is used as a unit of measurement to extract a 2D×2D size ROI image centring on the palm area. This method results in ROI images of variable sizes as shown in **Figure 11.**

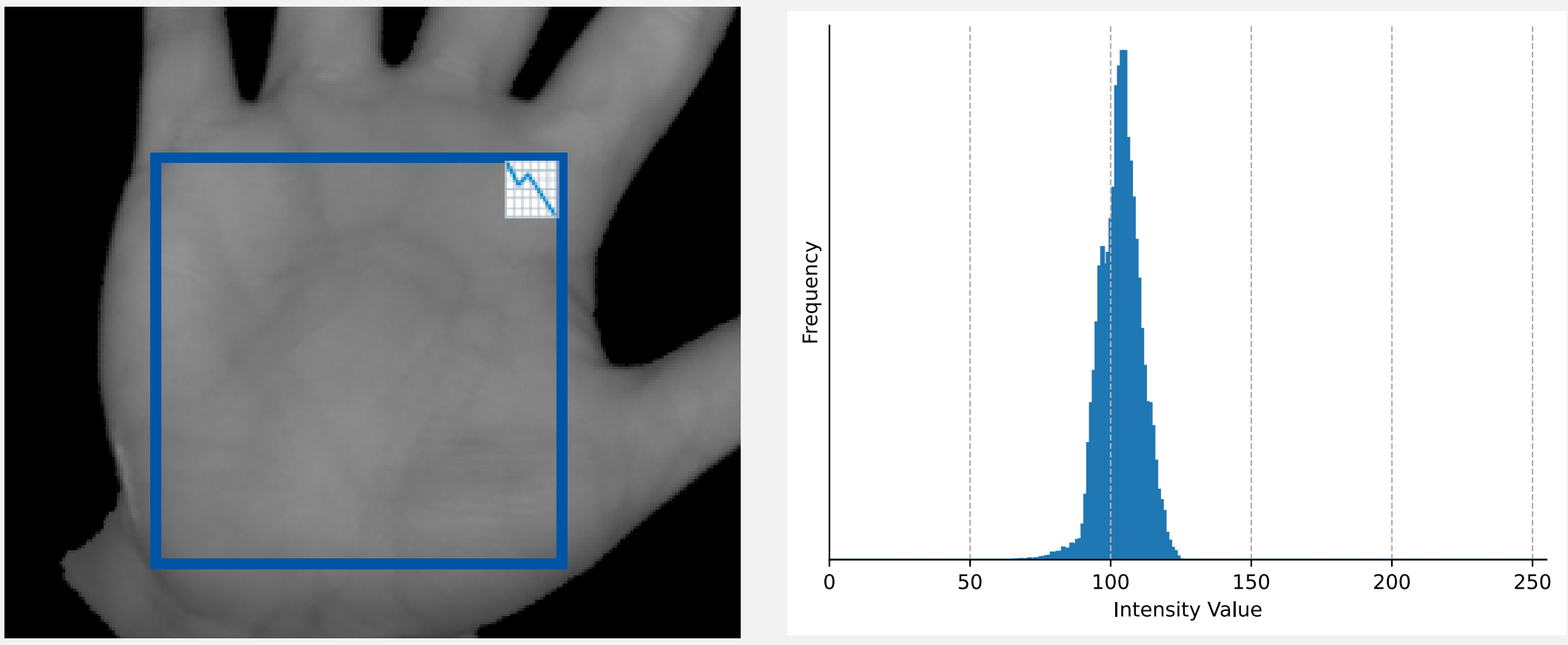

[a]: Sample image from the CASIA dataset [b]: Histogram of [a]

**Figure 12:** Sample image from the CASIA dataset and the corresponding histogram. ROI used to generate the histogram is marked with a blue rectangle.

An eroded copy of this matte was used as a mask to filter out the feature points residing closer to the edges. All the images were scaled down to 60% of their original size prior to any processing to speed up processing and reduce computational time.

## 5. Results and Discussion

In this section, first, the results of the ILACS-LGOT method are examined using visual outputs, histograms, and SIFT feature detection. All the visual comparisons are presented using the first right-hand image of subject '001' from the CASIA NIR dataset (001_r_850_01.jpg). The raw image and the histogram of the palmar area, highlighted with a blue square, are presented in **Figure 12**.

A comparison with the CLAHE method is also provided to highlight the differences and effectiveness of the ILACS-LGOT approach, followed by a visual examination and evaluation of SIFT feature matching and filtering. The section then compares the results with existing published research to provide a performance comparison against other image enhancement methods. Finally, it concludes with recommendations and observations based on the findings.

Determining the optimal falloff value is crucial for producing an acceptable final image. The optimal value identified for the standard deviation (also referred to as the falloff of the gradient) is $N/4.4$ for tile sizes between 8 and 32 pixels. $N$ is the length of an image tile as presented in ***Equation (2)***. Using lower falloff values will result in the sum of Gaussian weights on a pixel being lower than 1, rendering darker image tile edges, while higher falloff values will reveal the edges of the tiles. This will reproduce the blocky effect.

**Figure 13** showcases contrast enhancement results with the ILACS-LGOT method applied using four different tile sizes, when $N = \{16, 20, 28, 32\}$, along with corresponding SIFT features detected on these images and their histograms. It can be observed that when $N$ is smaller, the ILACS-LGOT method produces sharp detail enhancements. As $N$ increases, the image sharpness starts to gradually decrease, the darker areas start to become even darker, and shallow edges of the image tiles could appear for very large values. The corresponding histograms indicate that the intensity distribution remains closely the same for all the tile sizes, and most of the SIFT features are detected towards the central area of the palm.

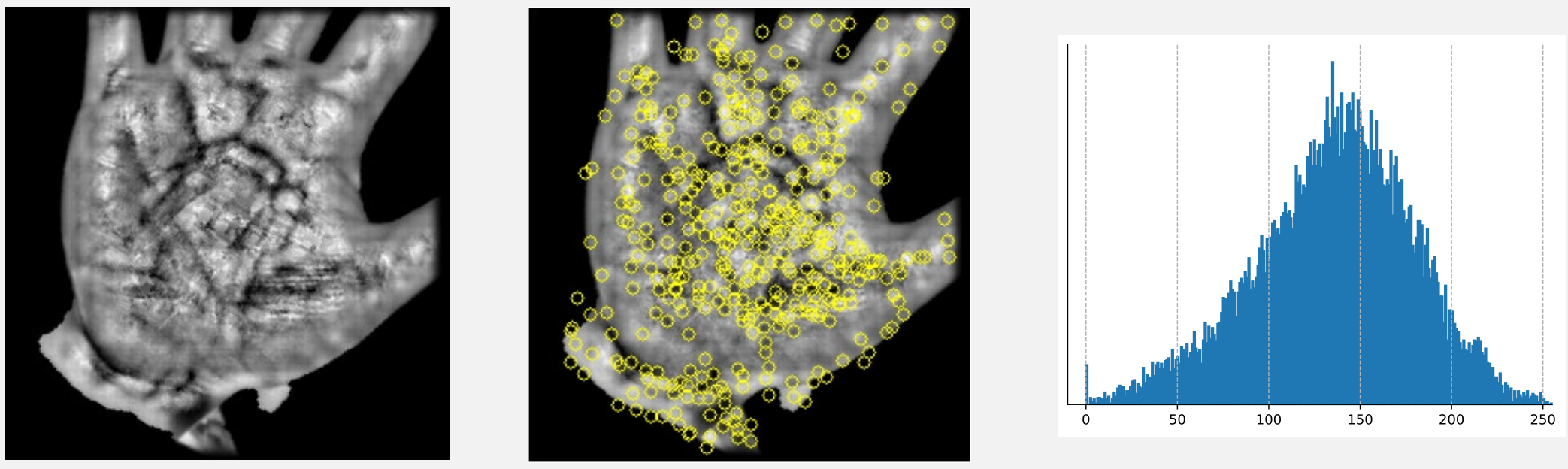

[a]: $N = 16$, Maximum set limit of 500 feature points.

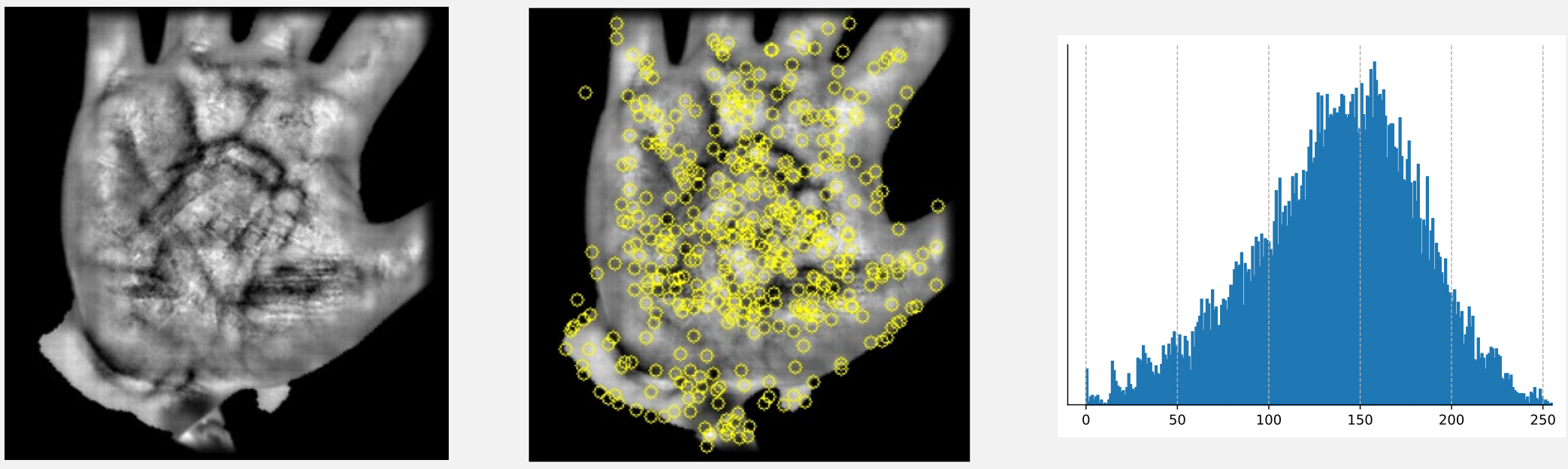

[b]: $N = 20$, Maximum set limit of 500 feature points.

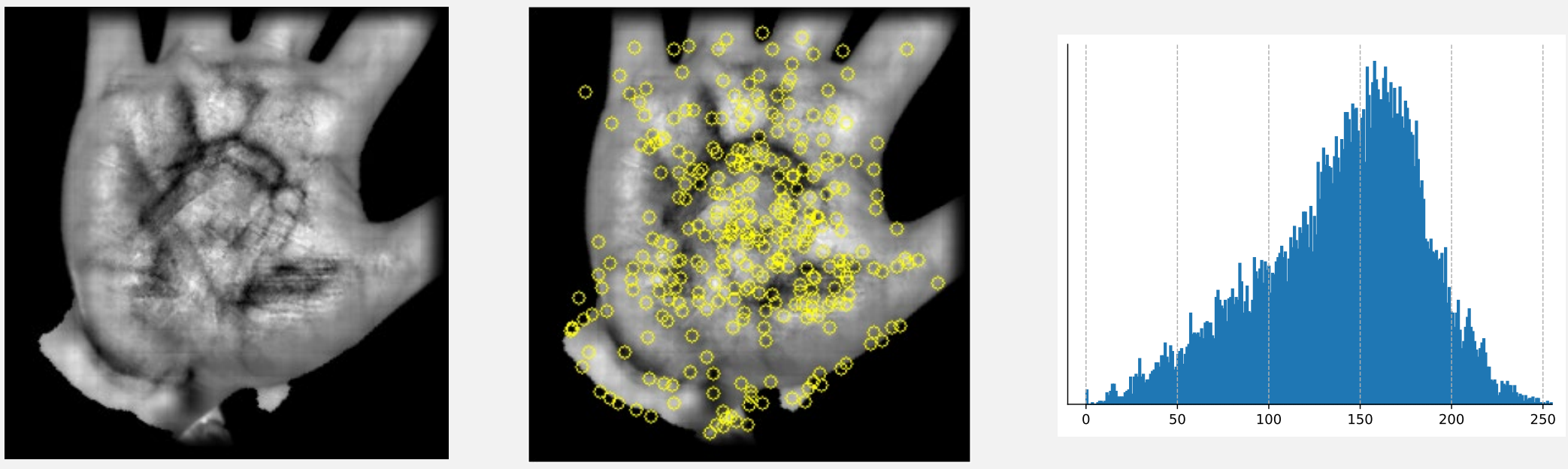

[c]: $N = 28$, 373 feature points.

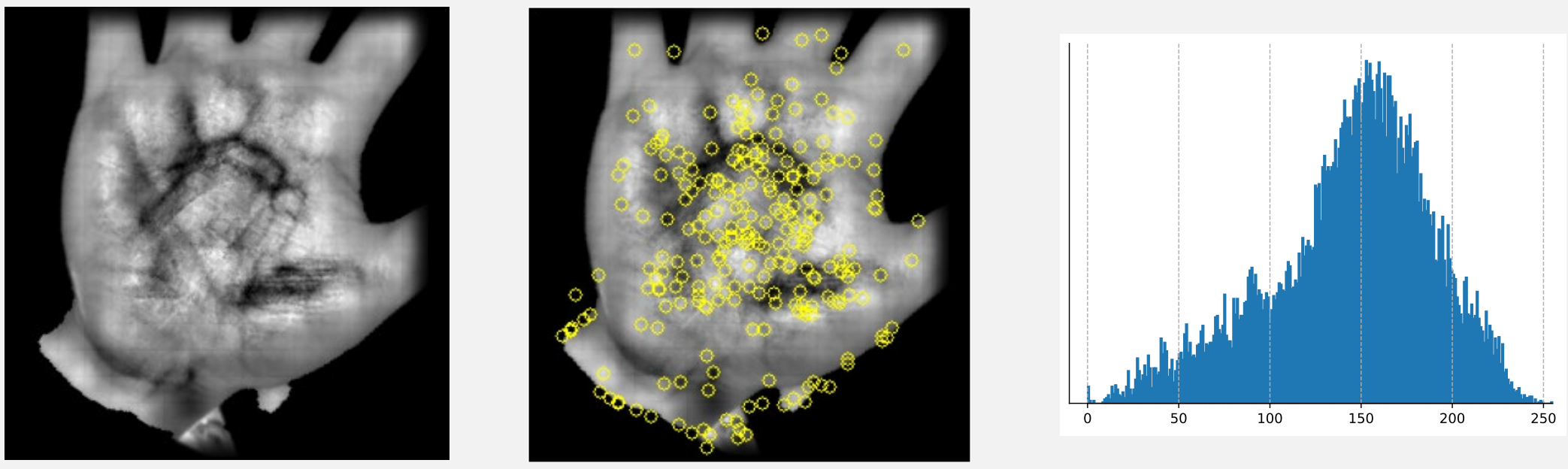

[d]: $N = 32$, 290 feature points.

**Figure 13:** ILACS-LGOT images and corresponding SIFT feature detection and the histograms, where $N = \{16, 20, 28, 32\}$. When $N$, (the tile size) is increased, image sharpness decreases, and the number of detected SIFT keypoints are decreased. The histogram shape remained relatively same across different tile sizes. The maximum number of SIFT points were set to 500.

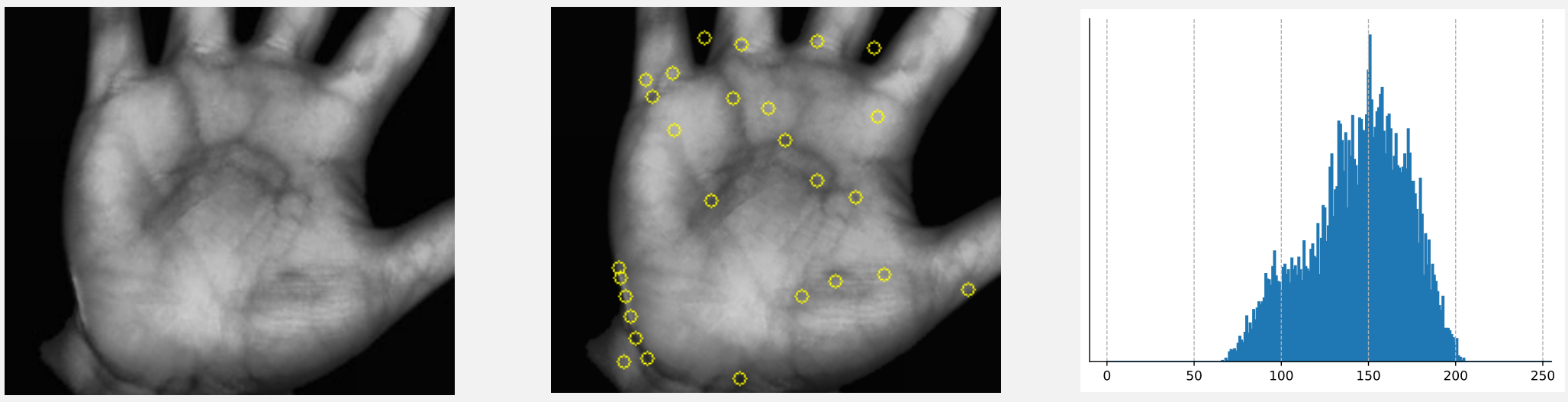

[a]: CLAHE, Tile-3, Clip-4, 29 feature points.

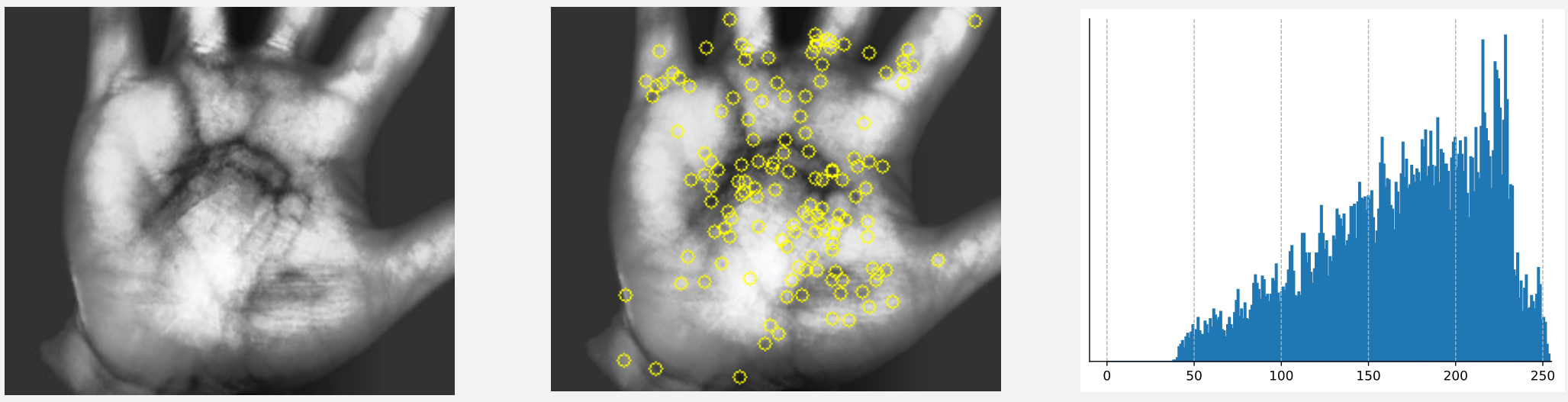

[b]: CLAHE, Tile-3, Clip-50, 165 feature points.

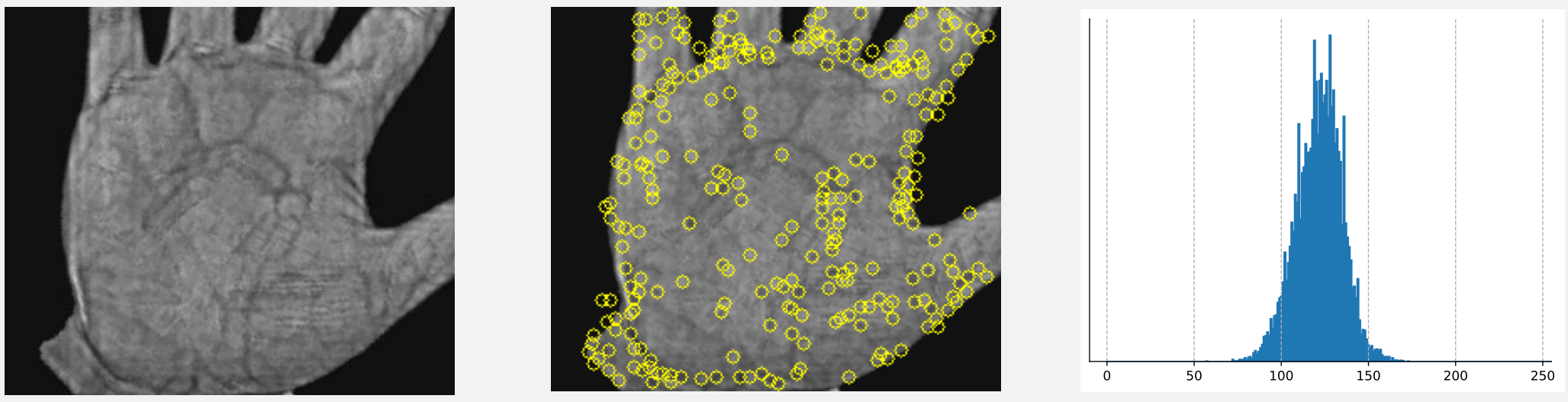

[c]: CLAHE, Tile-50, Clip-4, 316 feature points.

**Figure 14:** Comparison of SIFT feature point detection on CLAHE enhanced images with various parameter values.

## 5.1. Comparison with CLAHE

**Figure 14** examines CLAHE with various combinations of tile counts and clipping values, along with their corresponding histograms. These results are provided for comparison to demonstrate the superior performance of the ILACS-LGOT methods.

When using higher tile counts, CLAHE produces an undesirable grid or line pattern (**Figure 14** [c]), rendering the images unsuitable for palm-vein identification using feature points. This effect is highly noticeable when using a higher clip value. Conversely, when increasing the clip limit with a lower tile count, the finer details in the image are clipped and destroyed (**Figure 14** [b]).

Another observation of CLAHE is that images remain in low contrast distribution when using lower clip values (refer to **Figure 15** [a] and [c]). CLAHE with a lower tile count produces an insignificant number of SIFT features (**Figure 14** [a]). However, with a higher tile count, more SIFT features are detected, but these are mostly towards the edges of the palm (**Figure 14** [c]), which will be discarded or filtered out by a palm-vein matching algorithm. Nevertheless, CLAHE can produce results similar to those of HE when using a higher clip limit with a lower tile count (**Figure 14** [b]).

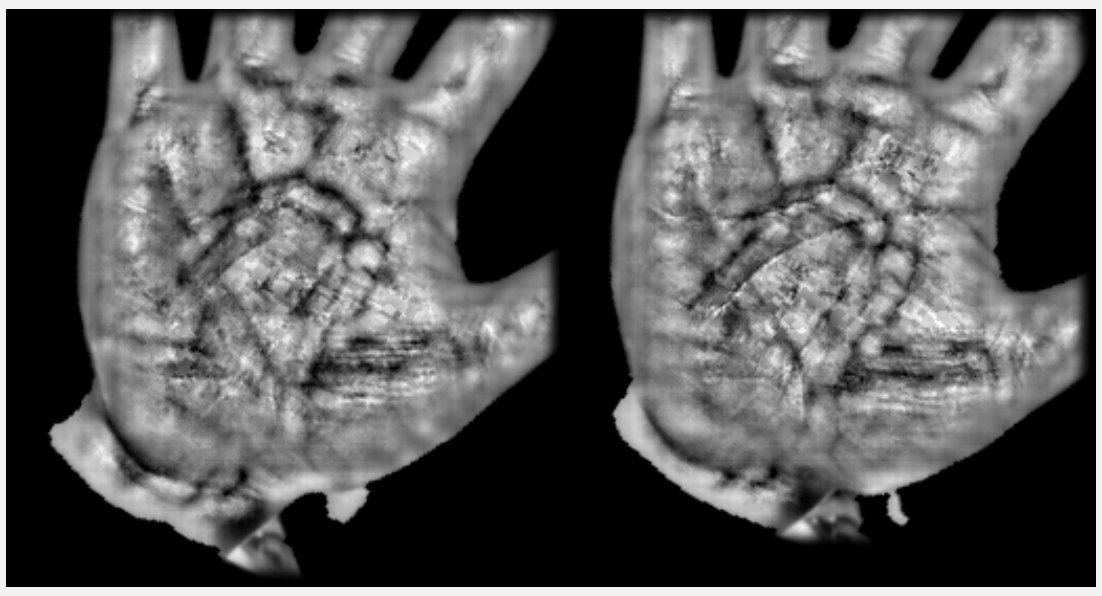

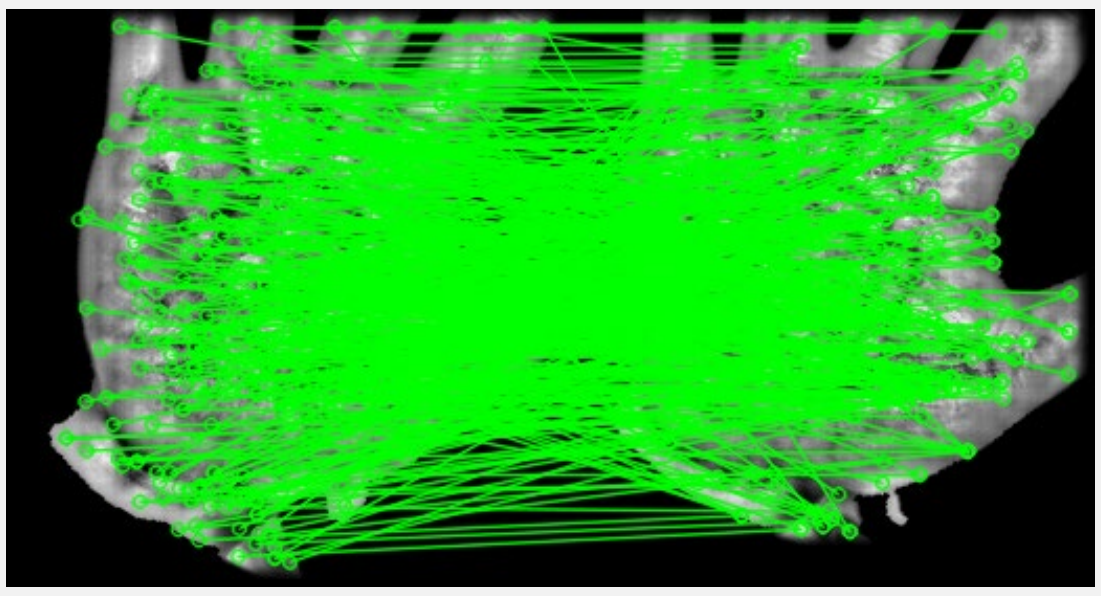

[a]: Pre-aligned ILACS-LGOT image pair.  [b]: ED Matching, no threshold applied.

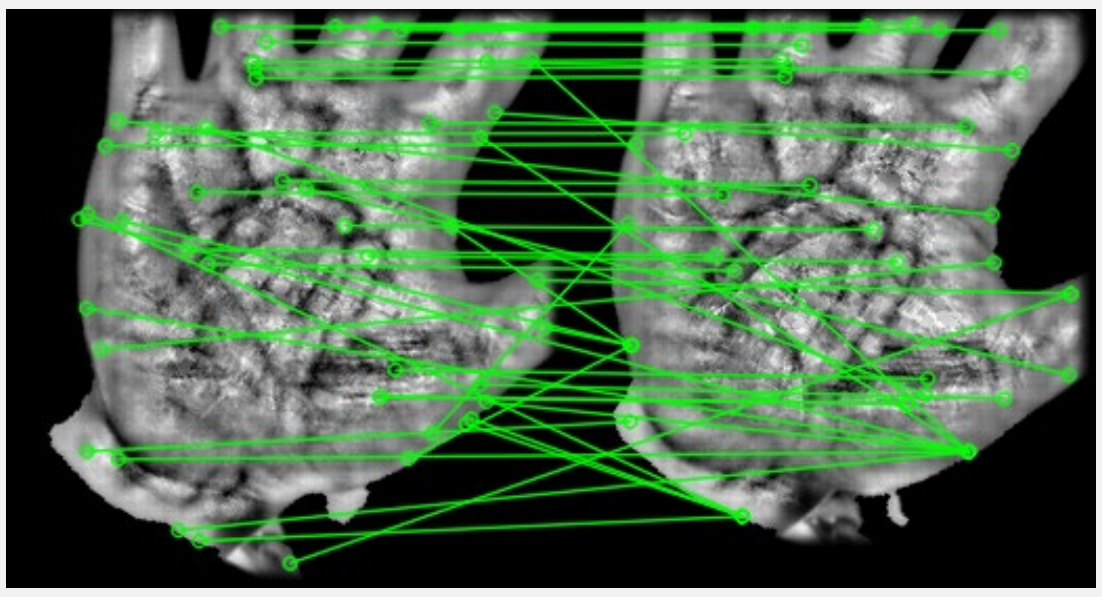

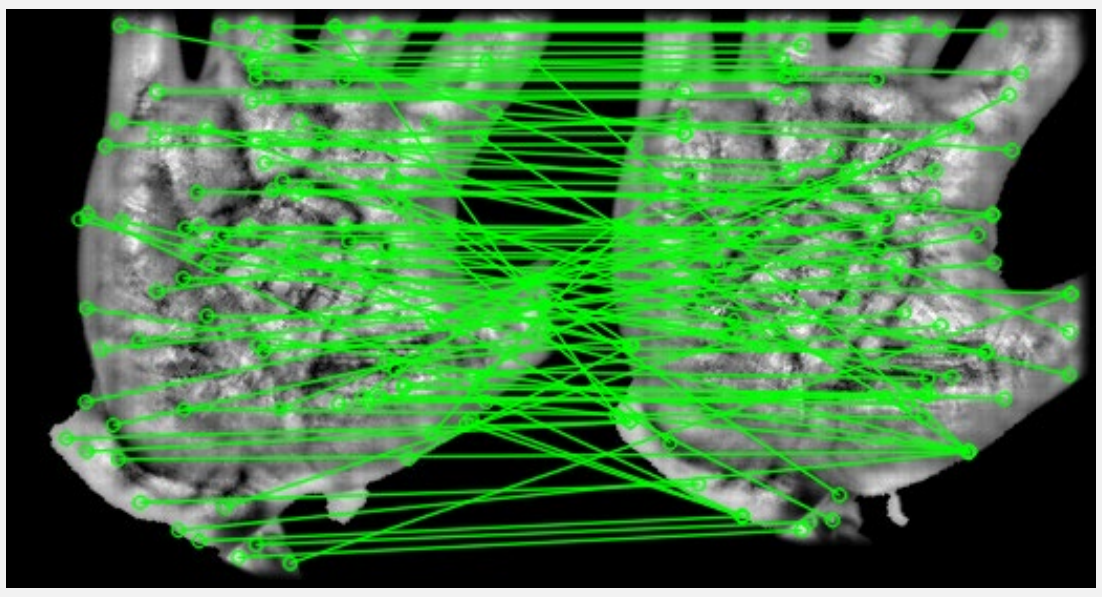

[c]: ED Matching, threshold applied - 250.  [d]: ED Matching, threshold applied - 300.

**Figure 15:** Closest-neighbour matching of SIFT features uses the Euclidean Distance (ED) as the distance measure. The keypoints near the edges were retained for demonstration purposes only.

## 5.2. SIFT Feature Matching and Parameters

This subsection will evaluate SIFT feature matching employing the ILACS-LGOT method and visually assess the parameters of widely used feature matching and filtering techniques. The primary feature-matching technique employed in the original SIFT implementation involves comparing the Euclidian Distance (ED) between feature descriptors [4].

**Figure 15** demonstrates the closest-neighbour SIFT matching of two palm images from the same subject, utilising ED. A threshold can be applied for closest-neighbour matching to filter out matches that exceed a set distance, as illustrated in subfigures [c] and [d] of **Figure 15**. However, this method is unreliable, as it may retain numerous incorrect matches.

To refine the matching process and reduce false positives, the distance ratio test (RT) introduced by Lowe [5] adds a layer of precision to the matching strategy. Instead of simply identifying the closest match by the smallest ED, the RT involves calculating the ratio of the distance to the nearest neighbour and the distance to the second-nearest neighbour.

A keypoint match is considered valid if the ratio of the smallest to the second-smallest ED is below a certain threshold, commonly set between 0.7–0.8 as suggested by Lowe. This 'Ratio Test' ensures that the selected match is distinctly closer than the next best match, thereby increasing the confidence in the robustness of the match. This method significantly reduces the likelihood of false matches by ensuring that the matches are not only close but also uniquely closer than any other potential matches.

The most commonly used closest-neighbour matching technique with SIFT feature matching is the k-nearest neighbours algorithm (KNN) [44].

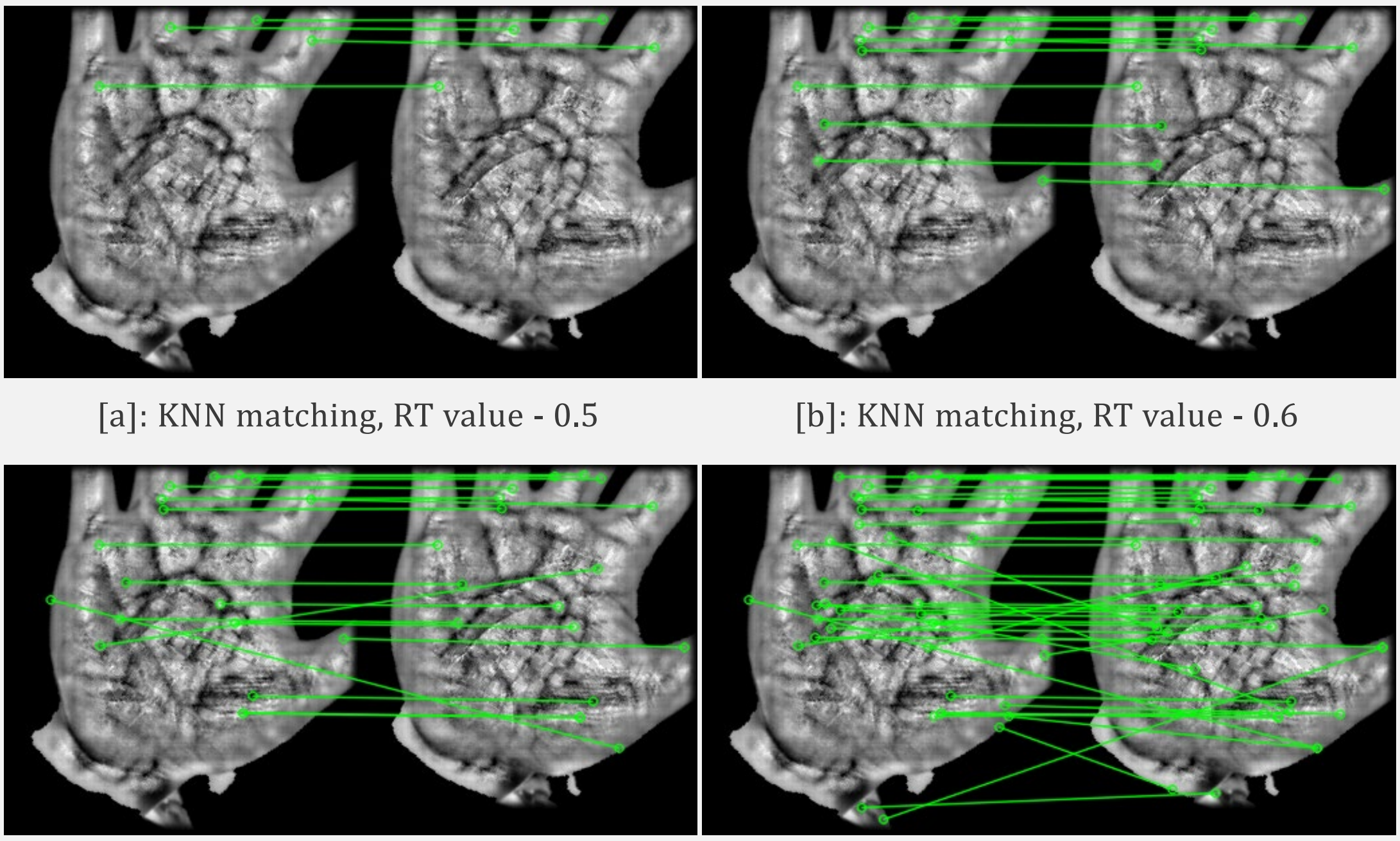

[a]: KNN matching, RT value - 0.5 [b]: KNN matching, RT value - 0.6

[c]: KNN matching, RT value - 0.7 [d]: KNN matching, RT value - 0.8

**Figure 16:** k-nearest neighbour (KNN) matching with the Ratio Test (RT) applied. Lower RT values can filter out many useful match pairs, while higher RT values may still preserve a considerable number of false match pairs.

**Figure 16** demonstrates second closest-neighbour matching with SIFT+KNN+RT, using various distance ratios.

RT can filter out almost all the false matches when using lower ratios. In the examples presented in subfigures [a] and [b] of **Figure 16**, a small number of the highest quality matches were retained when using RT values of 0.5 and 0.6, respectively.

Increasing the RT value, as presented in subfigures [c] and [d] of **Figure 16**, can start to filter in false matches, and the results gradually begin to appear similar to those of SIFT+ED. However, even with higher RT values, SIFT+KNN+RT can still preserve higher quality matches than SIFT+ED.

## 5.3. Performance Comparison with Existing Methods

To assess the proposed contrast enhancement method, several palm-vein image recognition systems that use SIFT and RootSIFT [45] features with the CASIA dataset were implemented. With SIFT features, ED [25] [43] [46] and ED+RANSAC filtering [20] were applied.

With RootSIFT features, the KNN [44], along with the RT [5] and the bidirectional matching methods presented in [43], were applied. These studies report EER of their systems. The pre-processing stage in these systems was replaced with the proposed ILACS-LGOT method using 16×16 image tiles, and its efficiency was verified using EER as a measure of performance.

The corresponding results are presented in **Table 1**. The proposed method reduces EER values when employing SIFT-based matching methods proposed by [43] and [20].

The recognition methods involving RANSAC filtering were computed with a significant increase in the computational time. The proposed contrast enhancement method can further reduce EER values when utilising RootSIFT-based matching.

**Table 1:** Performance comparison with existing palm-vein recognition systems.
The experimental results are presented using a template size of one unless otherwise noted.

| Recognition technique | EER % |
|---|---|
| Prewitt filter + SIFT (ED + bidirectional matching) [46] | 1.84 |
| **ILACS-LGOT + SIFT (ED + bidirectional matching)** | 7.77 |
| DoG-HE + SIFT (ED) [43] (Template size -3) | (Left hand) 2.87 |
| **ILACS-LGOT + SIFT (ED)** (Template size -3) | (Left hand) 1.48 |
| CLAHE + block stretch + SIFT (ED + RANSAC) [20] | 14.7 |
| **ILACS-LGOT + block stretch + SIFT (ED + RANSAC)** | 4.29 |
| GABOR + SIFT (ED) [25] | (L/R hands) 3.65/3.82 |
| CS-LBP + SIFT (ED) [25] | (L/R hands) 4.51/4.78 |
| ECS-LBP + SIFT (ED) [25] | (L/R hands) 3.12/3.25 |
| **ILACS-LGOT + SIFT (ED)** | (L/R hands) 2.75/2.88 |
| DoG-HE + RootSIFT (ED) [43] (Template size -3) | (Left hand) 2.16 |
| **ILACS-LGOT + RootSIFT (ED)** (Template size -3) | (Left hand) 1.27 |
| DoG-HE + RootSIFT (KNN + RT + Bidirectional matching) [43] (Template size -3) | (Left hand) 0.16 |
| **ILACS-LGOT + RootSIFT (KNN + RT + Bidirectional matching)** (Template size -3) | (Left hand) 0.4 |

In [43], Difference of Gaussians - Histogram Equalisation (DoG-HE) is utilised to enhance the image and extract SIFT and RootSIFT features. In this experiment, they used their own dataset and the 700nm, 850nm, and 940nm subsets of the CASIA multispectral dataset. In the experiment based on the 850nm CASIA subset, three images were used for a reference template, and three intra-class and six inter-class probe images were used. In [20], the image was first applied with CLAHE, then histogram stretching on image blocks was used to enhance the image prior to feature extraction. The SIFT features were first matched using ED, and then mismatches were estimated using the RANSAC algorithm.

In [25], ECS-LBP is applied prior to extracting SIFT features and matching using ED, and matching between left and right hands was separated. In the study in [46], a Prewitt high pass filter has been used to enhance the images before extracting SIFT and ORB features. Features are then matched using ED bidirectionally between a reference template and a probe image. The study further presents results with several multi-feature score fusion algorithms. However, the bidirectional matching method was implemented only with SIFT features. The recognition systems previously mentioned did not provide the parameter values used in their experiments, except in [43], where it was reported that a threshold of 0.8 was used for the RT.

Further, [25] presents separate EER values for left and right hands. They only match left hands with other left hands and right hands with other right hands, reducing the total number of inter-class matches by half. Experiments with separate left/right-hand matching indicated that reducing the total number of inter-class matches causes a drop in EER values.

**Table 3:** EER % values with SIFT on template sizes of 1–5.

| Template size | ED | ED + RANSAC | KNN+RT | KNN+RT+RANSAC |
|---|---|---|---|---|
| 1 | 3.33 | 4.29 | 2.87 | 4.97 |
| 2 | 1.68 | 3.21 | 1.72 | 3.02 |
| 3 | 1.72 | 2.9 | 1.66 | 1.87 |
| 4 | 0.53 | 1.3 | 0.61 | 1.12 |
| 5 | 0.31 | 0.64 | 0.28 | 0.57 |

**Table 2:** EER % values with RootSIFT on template sizes of 1–5.

| Template size | ED | ED + RANSAC | KNN+RT | KNN+RT+ RANSAC | KNN+RT+ Bidirectional |
|---|---|---|---|---|---|
| 1 | 1.98 | 5.66 | 2.53 | 5.39 | 2.5 |
| 2 | 1.26 | 3.83 | 1.7 | 2.58 | 0.89 |
| 3 | 1.42 | 3.8 | 1.16 | 2.22 | 0.65 |
| 4 | 0.59 | 1.55 | 0.51 | 1.16 | 0.31 |
| 5 | 0.34 | 1.29 | 0.26 | 0.62 | 0.1 |

Experiments with bidirectional matching methods on SIFT [46] and RootSIFT [43] resulted in higher EER values than the original research. Especially, the EER value difference is significantly high against [46]. However, EER values from experiments of the corresponding forward matching methods (SIFT+ED in [46] and SIFT+ED, RootSIFT+ED in [43]) are lower than those of the original research, which produced forward matching results using the same contrast enhancement method (DoG-HE) as the bidirectional method. This indicates the ILACS-LGOT method outperforms DoG-HE and all the other compared enhancement methods.

The proposed method was further tested using template sizes of 1–5 using ED, ED+RANSAC, KNN+RT, KNN+RT+RANSAC with both SIFT and RootSIFT. For RT, a threshold of 0.7 [5] was used for all the experiments. Additionally, with SIFT, ED+Bidirectional matching and with RootSIFT, KNN+RT+Bidirectional-matching was performed with an RT threshold of 0.8 [43].

These results are presented in **Table 2** and **Table 3**. As the template size increases, EER values are reduced for all methods except for ED matching when the template size is three. ED method performed slightly better with SIFT when using 4–5 samples per template. RANSAC based methods (ED+RANSAC and KNN+RT+RANSAC) performed better with SIFT features. However, ED alone produced lower EER values than when using ED+RANSAC method. All the other methods performed better with RootSIFT features. The lowest EER values are recorded when using RootSIFT features with KNN+RT+Bidirectional method (0.1%) followed by KNN+RT method (0.26%).

### 5.4 Observations and Recommendations

The ILACS-LGOT methods tend to amplify the noise presented within an image. It is recommended that the images be blurred prior to feature detection. Smaller tile sizes have a greater potential to amplify image noise, but they enhance the appearance of details in the local area. If the image size is increased, a similar visual effect can be achieved by proportionally increasing the tile size.

Scaling up the image to very high resolutions and applying the ILACS-LGOT method with smaller tile sizes produces visually interesting results, providing an enhanced view of the subtle details of the palm. However, these images are unsuitable for feature detection but can be useful in other applications.

The ILACS-LGOT prone to producing a blurred edge around the image. This can be mitigated by conditionally modifying the Gaussian mask-based weighting method to use a weighting of 1 if a given pixel is not present in multiple tiles. However, this will add another layer of computations and was not necessary for the application of palm-vein recognition. The feature points detected towards the edges can be discarded prior to matching.

The ILACS-LGOT method was developed using 8-bit images, which represent the dynamic range with values between 0 and 255. However, when dealing with floating-point images, where the dynamic range spans from 0 to 1, the flooring of the multiplying factor divided by the denominator may result in zero. To adapt these methods for floating-point images, the number of quantization levels supported by the image format (For instance, 10-bit image supports 1024) should be divided by the denominator, the result floored, and then divided again by the number of quantization levels to normalise between 0 and 1.

## 6. Conclusions

We proposed a novel palm-vein image enhancement method, termed ILACS-LGOT, designed to improve low-contrast palm vein images before feature extraction and subsequent processing, while retaining these subspace details.

The ILACS-LGOT method is based on adaptive contrast stretching, which subdivides an image into smaller tiles and adjusts the contrast within the available intensity range. To prevent the loss of details through clipping, we introduced an intensity-limited normalisation technique. The ILACS-LGOT method effectively enhances the contrast of fine details, irrespective of local illumination variations across different image regions. As a result, a higher number of SIFT features were detected in the central palm region.

We applied the widely used SIFT and RootSIFT recognition techniques to images enhanced with the ILACS-LGOT method. Additional experiments were conducted using a combination of methods with one to five registration images, and the results demonstrated that the proposed ILACS-LGOT method significantly outperforms existing image enhancement techniques.

The applicability of the ILACS-LGOT method extends beyond palm vein recognition, as it can be adapted for use in other closely related biometric modalities, such as finger vein and palmprint recognition. Furthermore, this approach can be further developed to enhance dark or low-contrast images in broader image processing applications

# References

[1] D. Palma and P. L. Montessoro, *Biometric-Based Human Recognition Systems: An Overview*. IntechOpen, 2022. doi: 10.5772/intechopen.101686.

[2] A. K. Jain, A. Ross, and S. Prabhakar, "An introduction to biometric recognition," *IEEE Trans. Circuits Syst. Video Technol.*, vol. 14, no. 1, pp. 4–20, Jan. 2004, doi: 10.1109/TCSVT.2003.818349.

[3] G. Wang and J. Wang, "SIFT Based Vein Recognition Models: Analysis and Improvement," *Comput. Math. Methods Med.*, vol. 2017, p. e2373818, Jun. 2017, doi: 10.1155/2017/2373818.

[4] D. G. Lowe, "Object recognition from local scale-invariant features," in *Proceedings of the Seventh IEEE International Conference on Computer Vision*, Sep. 1999, pp. 1150–1157 vol.2. doi: 10.1109/ICCV.1999.790410.

[5] D. G. Lowe, "Distinctive Image Features from Scale-Invariant Keypoints," *Int. J. Comput. Vis.*, vol. 60, no. 2, Art. no. 2, Nov. 2004, doi: 10.1023/B:VISI.0000029664.99615.94.

[6] Y. Zhou and A. Kumar, "Human Identification Using Palm-Vein Images," *IEEE Trans. Inf. Forensics Secur.*, vol. 6, no. 4, pp. 1259–1274, Dec. 2011, doi: 10.1109/TIFS.2011.2158423.

[7] K. Perera, F. Khelifi, and A. Belatreche, "A Novel Image Enhancement Method for Palm Vein Images," in *2022 8th International Conference on Control, Decision and Information Technologies (CoDIT)*, May 2022, pp. 1467–1472. doi: 10.1109/CoDIT55151.2022.9804034.

[8] P. Cancian, G. W. Di Donato, V. Rana, and M. D. Santambrogio, "An embedded Gabor-based palm vein recognition system," in *2017 IEEE EMBS International Conference on Biomedical & Health Informatics (BHI)*, Feb. 2017, pp. 405–408. doi: 10.1109/BHI.2017.7897291.

[9] N. A. M. Hayat, Z. M. Noh, N. M. Yatim, and S. A. Radzi, "Analysis of local binary pattern using uniform bins as palm vein pattern descriptor," *J. Phys. Conf. Ser.*, vol. 1502, p. 012043, Mar. 2020, doi: 10.1088/1742-6596/1502/1/012043.

[10] M. Wulandari, Basari, and D. Gunawan, "Evaluation of wavelet transform preprocessing with deep learning aimed at palm vein recognition application," presented at the THE 4TH BIOMEDICAL ENGINEERING'S RECENT PROGRESS IN BIOMATERIALS, DRUGS DEVELOPMENT, HEALTH, AND MEDICAL DEVICES: Proceedings of the International Symposium of Biomedical Engineering (ISBE) 2019, Padang, Indonesia, 2019, p. 050005. doi: 10.1063/1.5139378.

[11] A. Y. Pratiwi, W. T. A. Budi, and K. N. Ramadhani, "Identity recognition with palm vein feature using local binary pattern rotation Invariant," in *2016 4th International Conference on Information and Communication Technology (ICoICT)*, May 2016, pp. 1–6. doi: 10.1109/ICoICT.2016.7571952.

[12] G. R. Nayar, A. Bhaskar, L. Satheesh, P. S. Kumar, and Aneesh R.P, "Personal authentication using partial palmprint and palmvein images with image quality measures," in *2015 International Conference on Computing and Network Communications (CoCoNet)*, Dec. 2015, pp. 191–198. doi: 10.1109/CoCoNet.2015.7411186.

[13] Y.-Y. Fanjiang, C.-C. Lee, Y.-T. Du, and S.-J. Horng, "Palm Vein Recognition Based on Convolutional Neural Network," *Informatica*, vol. 32, no. 4, pp. 687–708, Jan. 2021, doi: 10.15388/21-INFOR462.

[14] S.-Y. Jhong *et al.*, "An Automated Biometric Identification System Using CNN-Based Palm Vein Recognition," in *2020 International Conference on Advanced Robotics and Intelligent Systems (ARIS)*, Aug. 2020, pp. 1–6. doi: 10.1109/ARIS50834.2020.9205778.

[15] M. H. Alshayeji, S. A. Al-Roomi, and S. Abed, "Efficient hand vein recognition using local keypoint descriptors and directional gradients," *Multimed. Tools Appl.*, vol. 81, no. 11, pp. 15687–15705, May 2022, doi: 10.1007/s11042-022-12608-6.

[16] R. Wang, G. Wang, Z. Chen, J. Liu, and Y. Shi, "An Improved Method of Identification Based on Thermal Palm Vein Image," in *Neural Information Processing*, T. Huang, Z. Zeng, C. Li, and C. S. Leung, Eds., in Lecture Notes in Computer Science. Berlin, Heidelberg: Springer, 2012, pp. 18–24. doi: 10.1007/978-3-642-34481-7_3.

[17] ManMohan *et al.*, "Palm Vein Recognition using Local Tetra Patterns," in *2015 4th International Work Conference on Bioinspired Intelligence (IWOBI)*, Jun. 2015, pp. 151–156. doi: 10.1109/IWOBI.2015.7160159.

[18] Y. Zhou and A. Kumar, "Human Identification Using Palm-Vein Images," *IEEE Trans. Inf. Forensics Secur.*, vol. 6, no. 4, Art. no. 4, Dec. 2011, doi: 10.1109/TIFS.2011.2158423.

[19] Y.-P. Lee, "Palm vein recognition based on a modified (2D)2 LDA," *Signal Image Video Process.*, vol. 9, no. 1, pp. 229–242, Jan. 2015, doi: 10.1007/s11760-013-0425-6.

[20] S. C. Soh, M. Z. Ibrahim, M. B. Yakno, and D. J. Mulvaney, "Palm Vein Recognition Using Scale Invariant Feature Transform with RANSAC Mismatching Removal," in *IT Convergence and Security 2017*, vol. 449, K. J. Kim, H. Kim, and N. Baek, Eds., in Lecture Notes in Electrical Engineering, vol. 449. , Singapore: Springer Singapore, 2018, pp. 202–209. doi: 10.1007/978-981-10-6451-7_25.

[21] J. Yang and B. Zhang, "Scattering Removal for Finger-Vein Image Enhancement," in *2011 International Conference on Hand-Based Biometrics*, Nov. 2011, pp. 1–5. doi: 10.1109/ICHB.2011.6094321.

[22] J. Yang and J. Yang, "Multi-Channel Gabor Filter Design for Finger-Vein Image Enhancement," in *2009*

*Fifth International Conference on Image and Graphics*, Sep. 2009, pp. 87–91. doi: 10.1109/ICIG.2009.170.

[23] W. Kim, "Visibility Restoration via Smoothing Speed for Vein Recognition," *IEICE Trans. Inf. Syst.*, vol. E102.D, no. 5, Art. no. 5, May 2019, doi: 10.1587/transinf.2018EDL8110.

[24] R. B. Trabelsi, A. Damak Masmoudi, and D. S. Masmoudi, "A bi-modal palmvein palmprint biometric human identification based on fusing new CDSDP features," in *2015 International Conference on Advances in Biomedical Engineering (ICABME)*, Sep. 2015, pp. 1–4. doi: 10.1109/ICABME.2015.7323236.

[25] H. T. Van, C. M. Duong, G. Van Vu, and T. H. Le, "Palm Vein Recognition Using Enhanced Symmetry Local Binary Pattern and SIFT Features," in *2019 19th International Symposium on Communications and Information Technologies (ISCIT)*, Sep. 2019, pp. 311–316. doi: 10.1109/ISCIT.2019.8905179.

[26] X. Yan, F. Deng, and W. Kang, "Palm Vein Recognition Based on Multi-algorithm and Score-Level Fusion," in *2014 Seventh International Symposium on Computational Intelligence and Design*, Dec. 2014, pp. 441–444. doi: 10.1109/ISCID.2014.93.

[27] W. Kang, Y. Liu, Q. Wu, and X. Yue, "Contact-Free Palm-Vein Recognition Based on Local Invariant Features," *PLOS ONE*, vol. 9, no. 5, Art. no. 5, May 2014, doi: 10.1371/journal.pone.0097548.

[28] I. T. Young, J. J. Gerbrands, and L. van Vliet, "Fundamentals of image processing," in *The Digital Signal Processing Handbook*, V. K. Madisetti and D. B. Williams, Eds., CRC Press, 1998, p. 51.01-51.91.

[29] B. Unal and A. Akoglu, "Resource efficient real-time processing of Contrast Limited Adaptive Histogram Equalization," in *2016 26th International Conference on Field Programmable Logic and Applications (FPL)*, Aug. 2016, pp. 1–8. doi: 10.1109/FPL.2016.7577362.

[30] K. Zuiderveld, "VIII.5. - Contrast Limited Adaptive Histogram Equalization," in *Graphics Gems*, P. S. Heckbert, Ed., Academic Press, 1994, pp. 474–485. doi: 10.1016/B978-0-12-336156-1.50061-6.

[31] S. M. Pizer *et al.*, "Adaptive histogram equalization and its variations," *Comput. Vis. Graph. Image Process.*, vol. 39, no. 3, pp. 355–368, Sep. 1987, doi: 10.1016/S0734-189X(87)80186-X.

[32] D. Vernon, *Machine vision: automated visual inspection and robot vision*. USA: Prentice-Hall, Inc., 1991.

[33] J. Y. A. Wang and E. H. Adelson, "Representing moving images with layers," *IEEE Trans. Image Process.*, vol. 3, no. 5, pp. 625–638, Sep. 1994, doi: 10.1109/83.334981.

[34] S. Dhanani and M. Parker, "7 - Alpha Blending," in *Digital Video Processing for Engineers*, S. Dhanani and M. Parker, Eds., Oxford: Newnes, 2013, pp. 49–52. doi: 10.1016/B978-0-12-415760-6.00007-6.

[35] T. Porter and T. Duff, "Compositing digital images," in *Proceedings of the 11th annual conference on Computer graphics and interactive techniques*, in SIGGRAPH '84. New York, NY, USA: Association for Computing Machinery, Jan. 1984, pp. 253–259. doi: 10.1145/800031.808606.

[36] Y. Hao, Z. Sun, and T. Tan, "Comparative studies on multispectral palm image fusion for biometrics," in *Proceedings of the 8th Asian conference on Computer vision - Volume Part II*, in ACCV'07. Berlin, Heidelberg: Springer-Verlag, Nov. 2007, pp. 12–21.

[37] M. Fischer, M. Rybnicek, and S. Tjoa, "A novel palm vein recognition approach based on Enhanced Local Gabor Binary Patterns Histogram Sequence," in *2012 19th International Conference on Systems, Signals and Image Processing (IWSSIP)*, Apr. 2012, pp. 429–432.

[38] W. L. Jhinn, M. G. K. Ong, L. S. Hoe, and T. Connie, "Contactless Palm Vein ROI Extraction using Convex Hull Algorithm," in *Computational Science and Technology*, R. Alfred, Y. Lim, A. A. A. Ibrahim, and P. Anthony, Eds., in Lecture Notes in Electrical Engineering. Singapore: Springer, 2019, pp. 25–35. doi: 10.1007/978-981-13-2622-6_3.

[39] J.-C. Lee, "A novel biometric system based on palm vein image," *Pattern Recognit. Lett.*, vol. 33, no. 12, Art. no. 12, Sep. 2012, doi: 10.1016/j.patrec.2012.04.007.

[40] P. Wang and D. Sun, "A research on palm vein recognition," in *2016 IEEE 13th International Conference on Signal Processing (ICSP)*, Nov. 2016, pp. 1347–1351. doi: 10.1109/ICSP.2016.7878046.

[41] W. Kang, Y. Liu, Q. Wu, and X. Yue, "Contact-Free Palm-Vein Recognition Based on Local Invariant Features," *PLOS ONE*, vol. 9, no. 5, Art. no. 5, May 2014, doi: 10.1371/journal.pone.0097548.

[42] R. S. Parihar and S. Jain, "A Robust Method to Recognize Palm Vein Using SIFT and SVM Classifier," in *Proceedings of International Conference on Sustainable Computing in Science, Technology and Management (SUSCOM)*, Amity University Rajasthan, Jaipur - India: Applied Computing eJournal, Feb. 2019. doi: 10.2139/ssrn.3356787.

[43] X. Yan, W. Kang, F. Deng, and Q. Wu, "Palm vein recognition based on multi-sampling and feature-level fusion," *Neurocomputing*, vol. 151, pp. 798–807, Mar. 2015, doi: 10.1016/j.neucom.2014.10.019.

[44] M. Muja and D. G. Lowe, "Scalable Nearest Neighbor Algorithms for High Dimensional Data," *IEEE Trans. Pattern Anal. Mach. Intell.*, vol. 36, no. 11, Art. no. 11, Nov. 2014, doi: 10.1109/TPAMI.2014.2321376.

[45] R. Arandjelović and A. Zisserman, "Three things everyone should know to improve object retrieval," in *2012 IEEE Conference on Computer Vision and Pattern Recognition*, Jun. 2012, pp. 2911–2918. doi: 10.1109/CVPR.2012.6248018.

[46] X. Yan, F. Deng, and W. Kang, "Palm Vein Recognition Based on Multi-algorithm and Score-Level Fusion," in *2014 Seventh International Symposium on Computational Intelligence and Design*, Dec. 2014, pp. 441–444. doi: 10.1109/ISCID.2014.93.